\documentclass[psfig]{aa}

\usepackage[authoryear]{natbib}
\usepackage{graphicx}
\usepackage{subfigure}
\usepackage{lscape}
\usepackage{morefloats}

\renewcommand{\subfigcapskip}{5pt}
\renewcommand{\subfigtopskip}{5pt}
\renewcommand{\subfigbottomskip}{5pt}

\begin{document}

\title{INFRARED SPECTROSCOPY OF FAINT 15 $\mu$m SOURCES IN THE HUBBLE DEEP FIELD 
SOUTH: FIRST HINTS AT THE PROPERTIES OF THE SOURCES OF THE IR BACKGROUND
\thanks{Based on observations with ISO, an ESA project with
instruments funded by ESA member states (especially the PI countries: France,
Germany, the Netherlands, and the United Kingdom) with the participation of ISAS
and NASA.} \thanks{Based on observations collected at the European Southern
Observatory, Chile, ESO No 63.O-0022, 65-000 67-000}}

\author{A. Franceschini\inst{1}, S. Berta\inst{1}, D.~Rigopoulou\inst{2}, 
H. Aussel\inst{3,4}, C.J. Cesarsky\inst{5}, D. Elbaz\inst{6}, R. Genzel\inst{2}, 
E. Moy\inst{2}, S. Oliver\inst{7}, M. Rowan-Robinson\inst{8}, P.P. Van der Werf\inst{9} 
}
\institute{$^1$ Dipartimento di Astronomia, Vicolo Osservatorio 5, I-35122 Padova,
Italy, E-mail: franceschini@pd.astro.it \\
$^2$ Max Planck Institute fuer Extraterrestrische Physik, Garching bei
Muenchen, Germany\\
$^3$ Osservatorio Astronomico di Padova, Italy \\
$^4$ Institute for Astronomy, 2680 Woodlawn Drive, Honolulu, Hawaii 96822, USA \\
$^5$ European Southern Observatory, Karl-Schwarzschild-str. 2,  85740 Garching, Germany\\ 
$^6$ Centre d'Etudes de Saclay, Service d'Astrophysique, Orme des Merisiers, 91191 Gif-sur-Yvette, France\\
$^7$ Astronomy Center, CPES, University of Sussex, Falmer, Brighton BN1 9QJ, UK\\
$^8$ ICSTM,  Astrophysics Group, Blackett Laboratory, Prince Consort Rd., London, 
SW2 1BZ, U.K.\\
$^9$ Leiden Observatory, P.O. Box 9513, 2300 RA, Leiden, The Netherlands. \\
}

\date{Received July 15, 2002/ Accepted .....}

\titlerunning{IR Spectroscopy of Faint ISO Galaxies}
\authorrunning{Franceschini, Berta, Rigopoulou, et al. }

\abstract{
We present a spectroscopic analysis of a
sample of 21 galaxies with z = 0.2--1.5 drawn from a 25 square arcmin ultra-deep ISOCAM 
survey at $\lambda_{eff}=15\ \mu$m centered in the WFPC-2 Hubble Deep Field South.
Near-infrared spectra are reported for 18 ISO sources, carried out
with ISAAC on the VLT, aimed at detecting the redshifted H$_\alpha+${\sc [Nii]}.
Additional optical data come from the ESO VLT/FORS2 and NTT/EMMI,
primarily targeting {\sc [Oii]}, {\sc [Oiii]} and H$_\beta$ for further physical insight.
Although not numerous in terms of areal density in the sky, this population of very 
luminous IR sources has been recently found to be responsible for a substantial 
fraction of the extragalactic background light energy density.
H$_\alpha$ line emission is detected in virtually all the observed 
objects down to a flux limit of 7$\times$10$^{-17}$ erg cm$^{-2}$ s$^{-1}$ 
(corresponding to L$_{H_\alpha} > 10^{41}$ erg s$^{-1}$ at z = 0.6 for H$_{\rm o}$ = 65, 
$\Omega_\Lambda=0.7$ and $\Omega_m$ = 0.3). 
Our analysis (including emission line, morphology, and SED properties) shows clear
evidence for AGN activity in only two of these sources: one type-I 
(with broadened H$_\alpha$ at z=1.57) and one type-II quasars (with inverted 
{\sc [Nii]}/H$_\alpha$ ratio at z=1.39), while we suspect the presence of an AGN
in two further sources (an Ultra-Luminous IR Galaxy, ULIRG, at z=1.27 and a luminous 
galaxy at z=0.69).
The H$_\alpha$ luminosities indicate star formation rates (SFR) in the remaining
sources between 0.5 and 20 M$_{\odot}/yr$, assuming a Salpeter IMF between 0.1 
and 100 M$_{\odot}$ and without extinction corrections. 
We find good correlations between the mid-IR, the radio and H$_\alpha$ luminosities, 
confirming the mid-IR light as a good tracer of star formation (while the SFR 
based on H$_\alpha$ flux show some large scatter 
and offset, which are still to be understood).
We have estimated the baryonic masses in stars with a newly-developed tool fitting 
the overall optical-IR continuum, and found that the host galaxies of ISO sources 
are massive members of groups with typically high rates of SF ($SFR\sim 10$ to 300
$M_\odot/yr$).
We have finally compared this ongoing SF activity with the already formed stellar
masses to estimate the timescales t$_{SF}$ for the stellar build-up, which turn-out
to be widely spread in these objects between 0.1 Gyrs to more than 10 Gyr.
The faint ISOCAM galaxies appear to form a composite population, including
moderately active but very massive spiral-like galaxies, and very luminous 
ongoing starbursts, in a continuous sequence. 
From the observed t$_{SF}$ and assuming typical starburst timescales, 
we infer that, with few exceptions, only a fraction of the galactic stars can be
formed in any single starburst event, while several of such episodes during a
protracted SF history are required for the whole galactic build-up.

}

\maketitle


\section{INTRODUCTION}

New populations of faint high-redshift sources have been recently discovered
by deep mid- and far-IR surveys with the Infrared Space Observatory (e.g.
Genzel \& Cesarsky 2000, Aussel et al. 1999, Altieri et al. 1999, 
Elbaz et al. 1999, Rowan-Robinson et al. 1997) and by
large millimetric telescopes at longer wavelengths (e.g. Smail et al. 1997, Hughes 
et al. 1998).

These sources display various distinct features compared with other optically 
selected galaxy populations. They are very luminous on average
($L_{bol}\geq 10^{11}\ L_\odot$, Elbaz et al. 2002), with the bulk of their 
emission coming out in the far-IR, in a similar way as the IRAS-selected galaxies
include the most luminous systems in the local universe. On the contrary, their areal
density (a few sources/square arcmin at the faintest limits detectable by ISO)
is much lower than found for faint blue galaxies in the optical (e.g. Ellis 1997).

Another remarkable property of the faint IR-selected sources is to display extremely 
high rates of evolution with redshift, exceeding those measured for galaxies at 
other wavelengths and comparable or larger than the evolution rates observed for 
quasars (Hughes et al. 1998, Barger et al. 1998, Elbaz et al 1999, 2002, Blain et 
al. 1999, Franceschini et al. 2001).     This fast evolution of the IR sources 
implies that dust-obscuration, if moderately important in local galaxies
where less than 50\% on average of the optical-UV emission is absorbed, has strongly 
affected instead the past active phases of galaxy evolution.

Franceschini et al. (2001) and Elbaz et al. (2002) have matched the statistical and
IR-spectral properties of the faint ISO sources detected at $\lambda_{eff}=15
\ \mu$m with the spectral intensity of the
recently discovered Cosmic IR Background (CIRB, see Hauser et al. 1998, Puget et al. 
1996). The CIRB appears to contain a large fraction (up to $\sim 70\%$, though this
number is made uncertain by the optical-UV background intensity) of the total 
extragalactic background energy density from radio to X-rays, Cosmic Microwave
Background excluded.
These analyses have found that, due to their high luminosities and moderate
redshifts ($z\simeq 0.5$ to 1.3), the faint 15 $\mu$m sources include the main 
contributors to the CIRB. This is a robust conclusion, based on the observed shape of 
the 15 $\mu$m counts, see Elbaz et al. 1999, and only assuming for these sources a 
typical IR galaxy SED.    Then a large fraction of stars in present-day galaxies, 
or alternatively the bulk of degenerate baryons contained in nuclear supermassive 
BH's, have formed during IR-luminous dust-extinguished evolutionary phases. 

The deep diffraction-limited observations at 15 $\mu$m with ISO 
provide quite an effective way of probing this dust-obscured 
high-redshift activity.   The numerous source samples detected in this way 
offer an important advantage over longer wavelength observations to allow easy 
optical identification, thanks to the relatively small error-box (4.6$''$ PSF,
Okumura 1998) and the moderate redshifts and faintness of the optical
counterparts. 

We report here on optical and near-IR spectroscopic follow-up of a representative 
subset of the faint ISO population selected from a region centered in the HDF South.
Apart from measuring the redshift, motivation for our observations was to clarify 
the nature of these objects: 
the main open issues were to assess the presence of energetically
dominant AGNs as power sources, and to estimate the main physical parameters
of the starburst and normal galaxy population, like the Star-Formation Rate (SFR),
the extinction, and the stellar mass.
Section 2 describes the IR-selected sample and the observations. 
Section 3 analyses the source properties based on the emission lines, while
Section 4 those of the SEDs and continuum emission. 
Section 5 compares optical, near-IR, far-IR and radio indicators of SF and discusses 
the level of activity in the IR-selected galaxies. 
Section 6 summarizes our conclusions.

All quantities are computed assuming a universal geometry with $H_0=65$ km s$^{-1}$ 
Mpc$^{-1}$, $\Omega_m=0.3$, $\Omega_\Lambda=0.7$.
We indicate with the symbol $S_{15}$ the flux density in Jy at 
15 $\mu m$ (and similarly for other wavelengths).

\section{THE SAMPLE AND THE OBSERVATIONS}

\begin{figure*}[!ht]
\vskip 16cm
\caption{Image of the HDF--S region with indicated the LW3 15 $\mu$m sources
(source labels refer to the catalogue by Aussel et al. 2003, see Tables \ref{tab:photometry2}
and following). Of the 86 objects indicated in this map, 63 belong to the complete
sample with $S_{15}>90 \mu$Jy, while the other are fainter than this limit.
The map is a collage by \cite{hook1999} including WFPC-2 F814 images in the Flanking
Fields. The map scale is 4.7 arcminutes on a side. 
Contours represent the LW3 image depth, increasing towards the center.  }
\label{fig:map}
\end{figure*}

\subsection{Sample selection}

The Hubble Deep Field South was observed between October 17 and November 29, 1997,
with the array camera ISOCAM onboard the Infrared Space Observatory,
as part of the ELAIS collaboration (Oliver et al. 2000).  The
observations were carried out with two broad-band filters, LW2 (5-8.5 $\mu$m, 
$\lambda_{eff}=6.75\ \mu$m) and LW3 (12-18 $\mu$m, $\lambda_{eff}=15\ \mu$m). 
This followed a previous similar observing campaign on the HDF-North
(Oliver et al. 1997), but adopted an improved observing strategy, particularly with LW2.
The deep ISOCAM images with the two filters, obtained as repeated raster scans to improve
the flat-field accuracy and time-redundancy, covered the same area centered on the 
HST WFPC-2 field.
All details on the observations can be found in Oliver et al. (2002).
We will consider in the following only the LW3 sample selected at 15 $\mu$m
(for these sources we will also make use of $6.75\ \mu$m fluxes or upper limits from the
LW2 observation).

Oliver et al. (2002) and Aussel et al. (2003) analyzed the ISOCAM data 
with two independent methods, accounting in detail for the time-varying signals
under the effect of cosmic ray impacts. 
In particular, the method adopted by Aussel et al. makes use of a wavelet analysis
of the combined spatial-temporal observable space (the PRETI method, see Stark et al. 
1999, Aussel et al. 1999).
The PRETI reduction has detected with LW3 63 sources brighter than
S$_{15 \mu m}$ = 90 $\mu$Jy (59 above 100 $\mu$Jy) over an area of 25 square arcminutes
(for comparison, 24 of these sources appear in the shallower list by Oliver et al. 
within the inner 19.6 square arcmins).  Detailed simulations have shown that above 
S$_{15 \mu m}$ = 100 $\mu$Jy the PRETI sample is complete and free of spurious sources.

\subsection{Source identification, target selection, photometric redshifts}
\label{par:2.2}

We have compared the ISOCAM source lists with those from the Deep ESO Imaging Survey
(EIS Deep), including optical imaging in UBVRI with ESO NTT/SUSI-2 down to
limiting magnitudes of U$_{AB}\sim 27$, B$_{AB}\sim 26.5$, V$_{AB}\sim 26$,
R$_{AB}\sim 26$, I$_{AB}\sim 25$ and near-infrared JHK observations down to
J$_{AB}\sim 25$, H$_{AB}\sim 24$, K$_{AB}\sim 24$ performed with NTT/SOFI 
(Da Costa et al. 1998).                  In addition we have used the optical
catalogues in uBVRI by \cite{teplitz1998}, which has 5$\sigma$ limiting
magnitudes of u$\sim$24.5,  B$\sim$26.1, V$\sim$25.5, R$\sim$25.4 and
I$\sim$24.
Unfortunately, the EIS Deep images  cover only a fraction ($\sim$70\%) of the
ISOCAM survey area. 


\begin{landscape}
\begin{table}
\centering
\footnotesize
\begin{tabular}{ l  c c r@{.}l  r@{.}l  r@{.}l  r@{.}l  r@{.}l  r@{.}l  r@{.}l r@{.}l  r@{.}l  c c c }
\hline
\hline
{\em Obj.} &  {\em R.A.} & {\em DEC.} & \multicolumn{2}{c}{\em U}  &  \multicolumn{2}{c}{\em B} &
 \multicolumn{2}{c}{\em V} & \multicolumn{2}{c}{\em R}  &
\multicolumn{2}{c}{\em I}  & \multicolumn{2}{c}{\em J} & \multicolumn{2}{c}{\em H} & \multicolumn{2}{c}{\em K} & \multicolumn{2}{c}{\em LW2} & {\em
$\sigma_{LW2}$} & {\em LW3} & {\em $\sigma_{LW3}$} \\
\hline
s14 & 22:32:41.52 &  -60:35:15.7& 22&11  & 21&74 & 20&82 & 20&28 & 20&04  &  19&80 & 19&67 & 19&46 & $<$0&143 & 0.143 & 0.239 & 0.044  \\
s16 & 22:32:42.89 &  -60:32:10.9& 22&57  & 22&38 & 22&00 & 21&35 & 21&07  &  20&81 & 20&73 & 20&42 & $<$0&038 & 0.038 & 0.123 & 0.039  \\
s19 & 22:32:43.51 &  -60:33:51.0& 20&37  & 20&09 & 20&23 & 19&96 & 19&74  &  19&78 & 19&42 & 19&53 & 0&195 & 0.029 & 0.288 & 0.050  \\
s20 & 22:32:44.11 &  -60:34:56.6& 21&79  & 21&49 & 20&58 & 19&96 & 19&50  &  18&99 & 18&68 & 18&54 & $<$0&053 & 0.053 & 0.164 & 0.041  \\
s23 & 22:32:45.59 &  -60:34:18.4& 24&90  & 24&11 & 22&17 & 21&42 & 20&78  &  19&84 & 19&26 & 18&94 & 0&118 & 0.023 & 0.749 & 0.097  \\
s25 & 22:32:45.81 &  -60:32:25.7& 24&14  & 23&43 & 22&73 & 21&88 & 21&37  &  20&27 & 19&80 & 19&58 & $<$0&023 & 0.023 & 0.473 & 0.071  \\
s27 & 22:32:47.70 &  -60:33:35.3& 22&22  & 21&87 & 20&97 & 20&07 & 19&46  &  18&86 & 18&39 & 18&17 & 0&059 & 0.021 & 0.387 & 0.061  \\
s28 & 22:32:47.61 &  -60:34:08.0& 23&10  & 22&83 & 22&26 & 21&64 & 21&18  &  20&80 & 20&36 & 20&24 & 0&046 & 0.021 & 0.172 & 0.042  \\
s38 & 22:32:53.13 &  -60:35:38.8& 27&42  & 25&88 & 25&46 & 24&86 & 23&72  &  22&25 & 21&67 & 21&53 & 0&123 & 0.023 & 0.518 & 0.075  \\
s39 & 22:32:53.06 &  -60:33:28.0& 25&03  & 24&63 & 24&28 & 23&74 & 23&09  &  21&49 & 20&95 & 20&82 & $<$0&015 & 0.015 & 0.226 & 0.043  \\
s40 & 22:32:52.91 &  -60:33:16.6& 25&98  & 25&20 & 24&78 & 24&15 & 23&49  &  21&99 & 21&34 & 21&09 & $<$0&006 & 0.006 & 0.119 & 0.038  \\
s43 & 22:32:53.75 &  -60:32:05.6& 26&02  & 25&92 & 25&21 & 24&63 & 23&61  &  22&48 & 21&92 & 21&62 & $<$0&015 & 0.015 & 0.095 & 0.035\\
s53 & 22:32:57.54 &  -60:33:05.5& 22&03  & 21&84 & 21&32 & 20&66 & 20&27  &  19&87 & 19&47 & 19&29 & 0&038 & 0.021 & 0.338 & 0.056  \\
s54 & 22:32:58.03 &  -60:32:04.2& 24&89$^{\diamond}$  & 23&82$^{\diamond}$ & 22&54$^{\diamond}$ & 21&39$^{\diamond}$ & 20&18$^{\diamond}$  &  \multicolumn{2}{c}{--} & \multicolumn{2}{c}{--} & \multicolumn{2}{c}{--} & $<$0&046 & 0.046 & 0.129 & 0.039  \\
s55 & 22:32:58.01 &  -60:32:33.8& 23&76  & 23&47 & 22&71 & 21&99 & 21&33 &  20&61 & 20&05 & 19&88 & $<$0&004 & 0.004 & 0.203 & 0.043  \\
s60 & 22:33:01.79 &  -60:34:12.9& 24&04  & 24&07 & 23&60 & 23&10 & 22&36 &  21&07 & 20&63 & 20&14 & $<$0&025 & 0.025 & 0.097 & 0.036  \\
s62 & 22:33:02.35 &  -60:35:25.3& 23&61  & 23&36 & 22&97 & 22&33 & 21&81 &  20&94 & 20&67 & 20&60 & $<$0&080 & 0.080 & 0.186 & 0.042  \\
s72 & 22:33:05.91 &  -60:34:36.3& 24&81$^{\diamond}$  & 23&70$^{\diamond}$ &  22&49$^{\diamond}$ & 21&34$^{\diamond}$ & 20&61$^{\diamond}$ &  19&89 & 19&27 & 18&94 & 0&079 & 0.021 & 0.370 & 0.059  \\
s73 & 22:33:06.17 &  -60:33:50.3& 19&33$^{\diamond}$  & 18&57$^{\diamond}$ &  17&73$^{\diamond}$ & 17&29$^{\diamond}$ & 16&86$^{\diamond}$ &  16&68 & 16&45 & 16&36 & 0&994 & 0.120 & 2.300 & 0.173  \\
s79 & 22:33:08.89 &  -60:34:34.3& 24&48$^{\diamond}$  & 24&38$^{\diamond}$ &  23&60$^{\diamond}$ & 22&99$^{\diamond}$ & 22&21$^{\diamond}$ &  21&63 & 21&11 & 20&78 & $<$0&049 & 0.049 & 0.186 & 0.042  \\
s82 & 22:33:12.42 &  -60:33:50.3& 23&04$^{\diamond}$  & 22&80$^{\diamond}$ &  22&02$^{\diamond}$ & 21&17$^{\diamond}$ & 20&36$^{\diamond}$ &  20&11 & 19&98 & 20&01 & 0&179 & 0.027 & 0.475 & 0.071  \\
\hline
\multicolumn{22}{l}{$^{\diamond}$: data from \cite{teplitz1998}}\\
\end{tabular}
\caption{Available photometric data for the 15 $\mu$m sources in the HDF--S observed
spectroscopically in the optical and near-infrared. Data are from the EIS Deep
survey, but in highlighted cases for which the Teplitz et al. (1998) photometry is used. 
All magnitudes are in the AB
system; ISOCAM LW2 and LW3 fluxes and uncertainties are reported in the last
four columns. In the case only an upper limit is available, it is specified.}
\label{tab:photometry2}
\end{table}
\end{landscape}

The spatial resolution of the ISOCAM images corresponds to a PSF of
4.6$''$ (Okumura 1998).
Within the ISOCAM errorbox, it turns out that there is almost invariably a galaxy
relatively bright in the red wavebands (I,J,H,K).
The procedures for optical identification are detailed in Mann et al. (2002)
(see also Aussel et al. 1999). 

Of the 63 sources in  the PRETI LW3 sample, 24 are outside the JHK EIS coverage
(10 of these are also outside the EIS optical imaging).

Of the other 39 sources, four are galactic stars, while for the remaining 35 objects
the optical/near-IR coverage of the galaxy SED is detailed enough to allow a precise 
estimate of the redshift through fits with spectro-photometric models.

We selected the targets for the VLT/ISAAC spectroscopic follow-up 
from the HDF South LW3 source list based on the following criteria: 
a) H$_\alpha$ should be in the wavelength range covered by the ISAAC gratings, 
b) a secure counterpart should exist in the I or the K band images.

We did not apply any selection based on redshifts or colours, except to ensure
that H$_\alpha$ was within the ISAAC spectral range. With these constraints,
our reference sample reduces to 25 galaxies with 15 $\mu$m flux densities
ranging between 95 and 800 $\mu$Jy. It is thus a representative sample of
the strongly evolving ISOCAM population near the peak of the
differential source counts (Elbaz et al. 1999). 

From these 25 sources we randomly selected 18 for the ISAAC follow up. For part
of these and for 3 additional objects we have optical spectroscopic data.

The source list, coordinates, and the avaliable optical, near-IR and mid-IR photometric
data are reported in Table \ref{tab:photometry2}. Stamp images from the F814 WFPC-2
maps for each sources are reported in Appendix \ref{par:sources}, together with 
plots of the optical-IR SEDs and notes on the individual sources.

\begin{figure}[!ht]
\centering\rotatebox{-90}{
\includegraphics[height=0.48\textwidth]{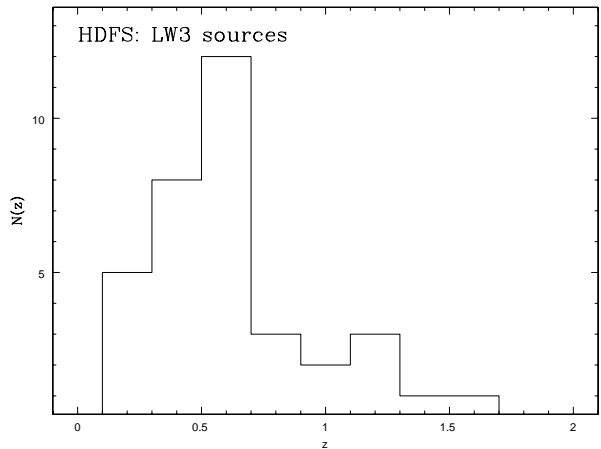} }
\caption{The redshift distribution of faint ISOCAM LW3 sources in HDF South
with optical identifications.
The objects range between z=0.2 and 1.6, mostly due to K-correction effects.
Redshifts are spectroscopic when available, otherwise photometric.}
\label{fig:z_distr}
\end{figure}

To set the ISAAC grating (Z, SZ, J, H) for H$_\alpha$ detections in our target sources, 
we used redshifts from optical spectra available for some of the objects at
z$<$0.7 (Dennefeld, private communication, see also Rigopoulou et al. 2000).   

For all other sources we used photometric redshift estimates based on fits of the
observed SEDs with synthetic galaxy spectra. In particular, we have used 
PEGASE (Fioc and Rocca-Volmerange 1997) and GRASIL (Silva et al. 1998) 
to construct grids of spectra as a function of galactic age,
considering two evolutionary sequences: one describing a spiral-like evolutionary
model with exponentially decreasing star-formation and long (5 Gyr) timescale,
the other reproducing a more typical evolution for ellipticals with short (1 Gyr) 
exponential timescale and star-formation truncated by a galactic wind.

\begin{table}[!ht]
\centering
\footnotesize
\begin{tabular}{c c c c}
\hline
\hline
Source&Run &          PA    &    Grating\\
\hline
s16   &     1999 &        -49.99 &    SZ\\
s19  &      2000 &         75.42 &    H\\
s23   &     1999 &         63.45 &    Z\\
s25$^\ast$ &      1999,2000&    65.45&    SZ\\
s27$^\ast$ &      1999,2000&    -43.85 &    SZ\\
s28  &      1999   &      -31.11 &    SZ\\
s38  &      1999  &       -77.41 &    H\\
s39  &      1999  &        29.45 &    H\\
s40 &       2000  &       -5.55  &    H \\
s43 &       2000 &        -56.05 &    J\\
s53 &       1999 &         67.22 &    SZ\\
s54  &      2000  &        10.25 &    J\\
s55$^\ast$ &      1999,2001 &   -56.05 &    J\\
s60  &      1999,2000 &   -77.41 &    H\\
s62  &      2000   &       64.15 &    J\\
s72  &      2001  &        -67.35 &    SZ\\
s79  &      2000  &        11.57  &   J\\
s82  &      2000  &        110.30 &   J\\
\hline
\multicolumn{4}{l}{$^\ast$: high-res data ($R_s\sim5000$) also exist.}\\
\end{tabular}
\caption{Summary of ISAAC/VLT observations. Slits have been rotated 
(anti--clockwise with respect to the North direction by the angle in column PA)
in order to include the most interesting features of the sources.}
\label{tab:obs_log}
\normalsize
\end{table}

Since the main spectral feature of the SED suited for the redshift estimates is the 
Balmer 4000 $\AA$ discontinuity, the effect of dust extinction on the redshift
estimate is modest.
The fit is automatically found from $\chi^2$ interpolation on a 2D grid of redshift
versus galactic age. 

After spectroscopic confirmation, our photometric redshift determinations turned out 
to be accurate to $\Delta$z$\simeq \pm$0.1. The fits based on the GRASIL and PEGASE
codes provided essentially the same results in terms of redshifts.
The distribution of redshifts for the ISOCAM LW3 sample in the HDF South is shown 
in Fig. \ref{fig:z_distr}.
The objects range between z=0.2 and 1.6, an interval 
mostly imposed by K-correction effects
(e.g. Franceschini et al. 2001): at $z\sim0.6-0.8$ the 7.7 $\mu$m PAH emission feature 
usually very intense in starburst galaxies falls within the LW3 filter.
The z distribution shows a strong peak at $z\simeq 0.6$ probably due to a cluster or a
large galaxy concentration in the HDF South. This overdensity is also apparent
in the analysis of photometric redshifts by Rudnick et al. (2001).
This z distribution is remarkably different from that of ISOCAM sources at 
similar depths in the HDF North (Aussel et al. 1999), particularly
in lacking the peak at z$\sim 0.8-1$.
The effect of cosmic variance is very prominent between the two fields.

\subsection{IR spectroscopic observations and data reduction}

We collected the IR spectra during three runs (September 1999, August 2000 and 2001)
using the infrared spectrograph ISAAC (Moorwood et al. 1998) on the ESO ANTU
telescope (formerly UT1), on Paranal, Chile.  The observing logs are sumarized
in Table \ref{tab:obs_log}.

Observations have been performed with the Low Resolution grating, providing a 
spectral resolution R$_{s}\sim$600 for a slit width of 1$''$ 
(the length is fixed to 2$'$). 
Four of our sample objects (S55, S27, and an interacting pair associated with source S25) 
were also observed during the 2000 and 2001 runs with the Medium-Resolution gratings 
(R$_{s}\sim$5000) (Rigopoulou et al. 2002).


\begin{table*}[!ht]
\centering
\footnotesize
\begin{tabular}{ l c c c    c c  c c  c c  c c  c c c c}
\hline
\hline
\multicolumn{4}{c}{Object}&&
\multicolumn{2}{c}{{\sc [Oii]}} &&
\multicolumn{2}{c}{H$\beta$} &&
\multicolumn{2}{c}{{\sc [Oiii]}} &&
\multicolumn{2}{c}{H$\alpha$(+{\sc [Nii]})}\\
\cline{1-4}
\cline{6-7}
\cline{9-10}
\cline{12-13}
\cline{15-16}
 \# & Instr. & $t_{exp}$ & z && S & $|EW|$ && S & $|EW|$ && S & $|EW|$ && S & $|EW|$\\
\multicolumn{15}{c}{}\\
s14  & EMMI & 5400s & 0.41 && $19.4$ & 26 && $9.55$ & 10 && $6.64$ & 7  &&       & \\
s16$^\dagger$  & ISAAC& 3720s & 0.62 &&        &    &&        &    &&        &    && $11.7$ & 45 \\
s19  & ISAAC& 5760s & 1.57 &&        &    &&        &    &&        &    && $157$ & 816 \\
s20  & EMMI & 5400s & 0.39 && $17.2$ & 37 && $7.30$ & 7  &&        &    &&       & \\
s23  & EMMI & 8100s & 0.46 && $13.7$ & 28 && $3.23$ & 9  && $6.04$ & 16 &&       & \\
s23$^\dagger$  & ISAAC& 3720s & 0.46 &&        &    &&        &    &&        &    && $18.6$ & 50 \\
s25  & EMMI & 5400s & 0.58 &&        &	  && $4.90$ & 18 &&        &    &&       & \\
s25  & FORS2& 17380s & 0.58  &&        &    && $5.64$ & 28 &&        &    &&       & \\
s25$^\dagger$  & ISAAC& 3720s & 0.58 &&        &    &&        &    &&        &    && $31.2$ & 110 \\
s27  & FORS2& 15340s & 0.58 &&        &    && $6.03$ &  4 &&        &    &&       & \\
s27  & FORS2& 18000s & 0.58 &&        &    && $7.39$ &  6 &&        &    &&       & \\
s27$^\dagger$  & ISAAC& 3720s & 0.58 &&        &    &&        &    &&        &    && $32.82$ & 47 \\
s28$^\dagger$  & ISAAC& 3720s & 0.58 &&        &    &&        &    &&        &    && $7.8$ & 47 \\
s38$^\dagger$  & ISAAC& 7400s & 1.39 &&        &    &&        &    &&        &    && $19.5$ & 35 \\
s39$^\dagger$  & ISAAC& 7400s & 1.27 &&        &    &&        &    &&        &    && $71.3$ & 67 \\
s40  & ISAAC& 3860s & 1.27 &&        &    &&        &    &&        &    && 13.1  & 67\\
s40  & ISAAC& 5760s & 1.27 &&        &    &&        &    &&        &    && 13.6  & 95\\
s43  & ISAAC& 4320s & 0.95 &&        &    &&        &    &&        &    && 38.6  & 13\\
s53(I)& EMMI & 7200s & 0.58 && $11.6$ & 39 && $12.0$ & 30 &&        &    &&       & \\
s53(I)& FORS2& 18000s &0.58 &&        &    && $16.7$ & 33 &&        &    &&       & \\
s53$^\dagger$  & ISAAC& 3720s &0.58 &&        &    &&        &    &&        &    && 60.8  & 70\\
s55    & FORS2& 18000s &0.76 &&        &    && $3.94$ & 28 &&        &    &&       & \\
s55	&ISAAC& 3840s  &0.76 &&		&	&&	&	&&	&	&& 30.7& 67 \\
s55$^\dagger$  & ISAAC& 3720s & 0.76 &&        &    &&        &    &&        &    && 24.1  & 40\\
s60$^\dagger$  & ISAAC& 7400s & 1.23 &&        &    &&        &    &&        &    && 27.3  & 44\\
s62$^\dagger$  & ISAAC& 3720s & 0.73 &&        &    &&        &    &&        &    && 25.4  & 62\\
s72 &	ISAAC &	1800s & 0.55	&&		&	&&	&	&&	&	&& 53.5&111 \\
s73    & FORS2& 18000s & 0.17 &&        &    &&        &    &&        &    && 224  & 32\\
s79    & ISAAC& 3840s & 0.74 &&        &    &&        &    &&        &    && 13.2  & 65\\
s82    & ISAAC& 3840s & 0.69 &&        &    &&        &    &&        &    && 33.2  & 26\\
\hline
\multicolumn{15}{l}{$^\dagger$: data from \cite{rigo2000}}
\end{tabular}
\caption{Summary of results of spectroscopic observations. We report the source name,  
the instrument, the exposure time, spectroscopic redshift, the measured fluxes 
(in units of $10^{-17}$ erg cm$^{-2}$ s$^{-1}$) and equivalent widths (EW, in \AA)
of {\sc [Oii]}$\lambda3727$, H$\beta$ ($\lambda$ 4861), {\sc [Oiii]}$\lambda \lambda 
4959, 5007$, and H$\alpha$ ($\lambda$ 6563).
H$\alpha$ fluxes and EW of H$\alpha$+{\sc [Nii]} are from \cite{rigo2000}. These
measured line fluxes are not corrected for aperture.
}
\label{tab:dati_spec}
\normalsize
\end{table*}

To maximize the observing efficiency, whenever possible the slit position included two 
target galaxies at any given orientation.  Most of the targets were first
acquired directly from a 1--2 min exposure in the H-band. In the case
of the very faint objects (H $\geq$ 20.0 mag), we blind-offset from a
brighter star in the HDF-S field.  Observations were made by nodding
the telescope $\pm$ 20$''$  along the slit  to facilitate sky
subtraction (always avoiding overlap of the two objects in the
slit). Individual exposures range from
120 (in H and J bands) to 300 (SZ band) seconds. 
During the 1999 and 2000 runs, sky conditions were excellent throughout the acquisition 
of the spectra, with seeing values typically in the range 0.4$''$--0.8$^
{\arcsec}$ and dipping down to 0.25$''$.  For each
filter, observations of spectroscopic standard stars were made in
order to flux calibrate the galaxy spectra.

\begin{figure*}[!ht]
\centering
\subfigure[HDF--S: s19.]{
\label{fig:spettro_s19H_isaac}
\rotatebox{-90}{
\includegraphics[height=0.48\textwidth]{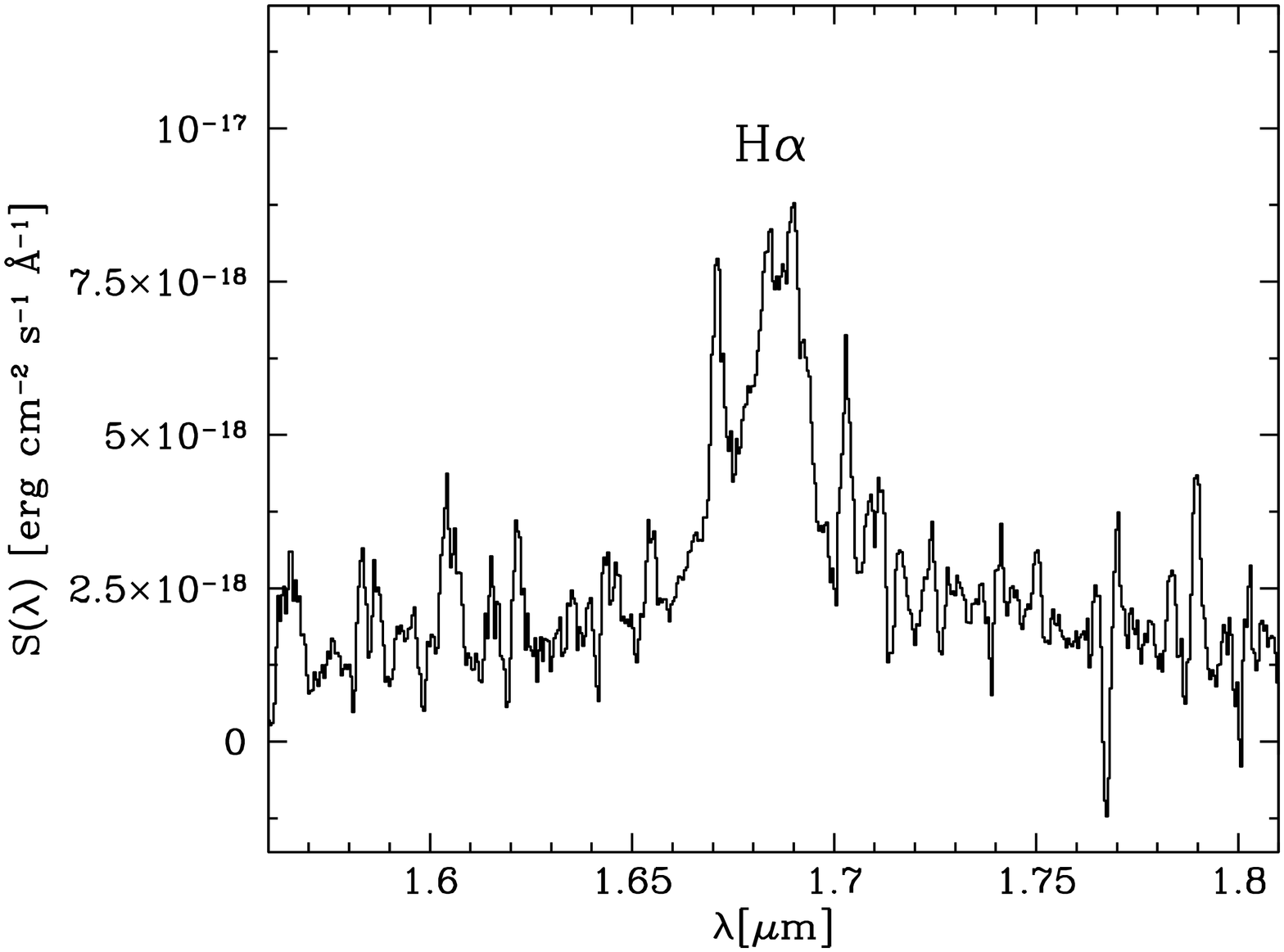}}}
\subfigure[HDF--S: s40.]{
\label{fig:spettro_s40H_isaac_19_08}
\rotatebox{-90}{
\includegraphics[height=0.48\textwidth]{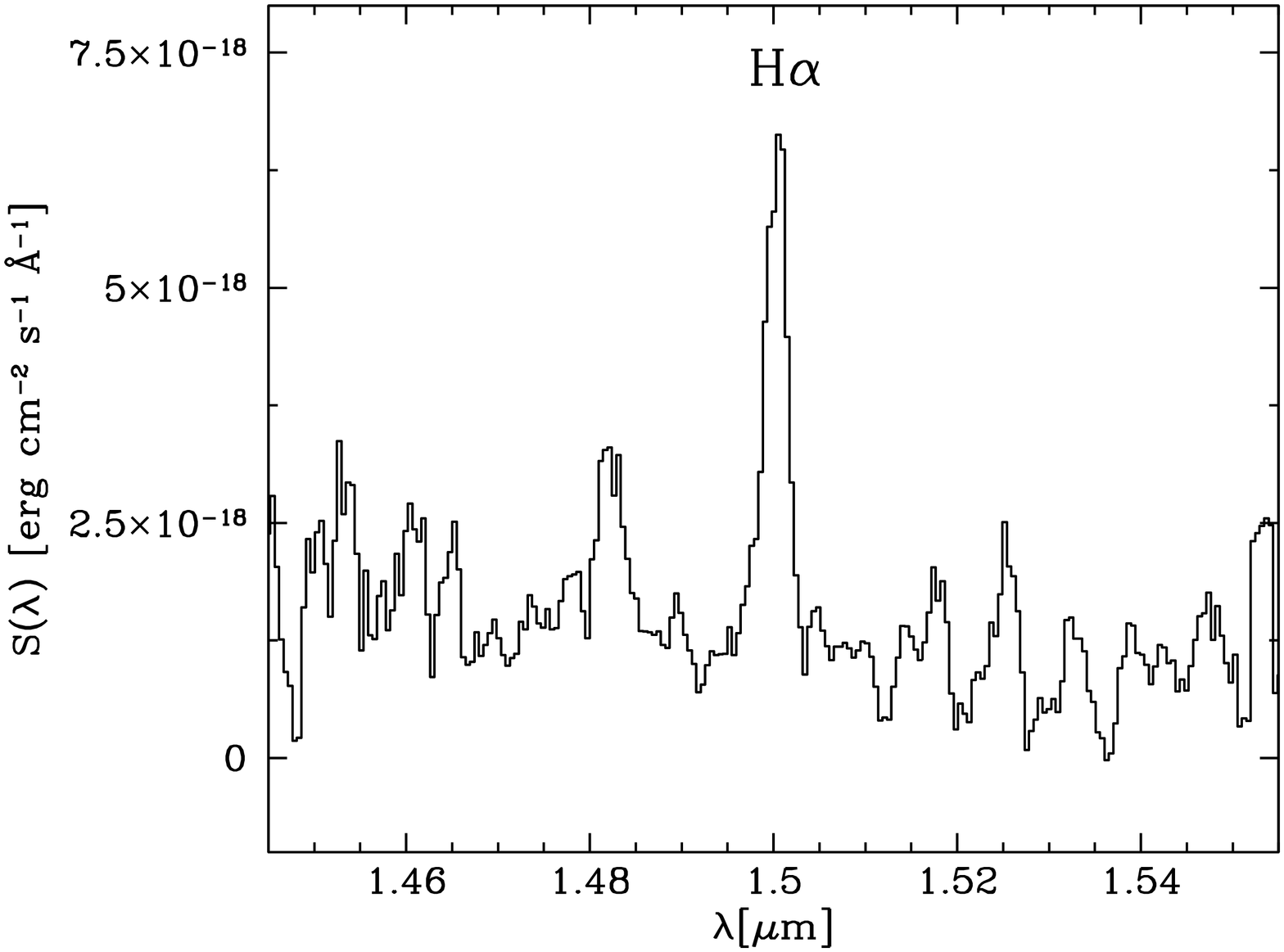}}}
\subfigure[HDF--S: s55.]{
\label{fig:spettro_s55J_isaac}
\rotatebox{-90}{
\includegraphics[height=0.48\textwidth]{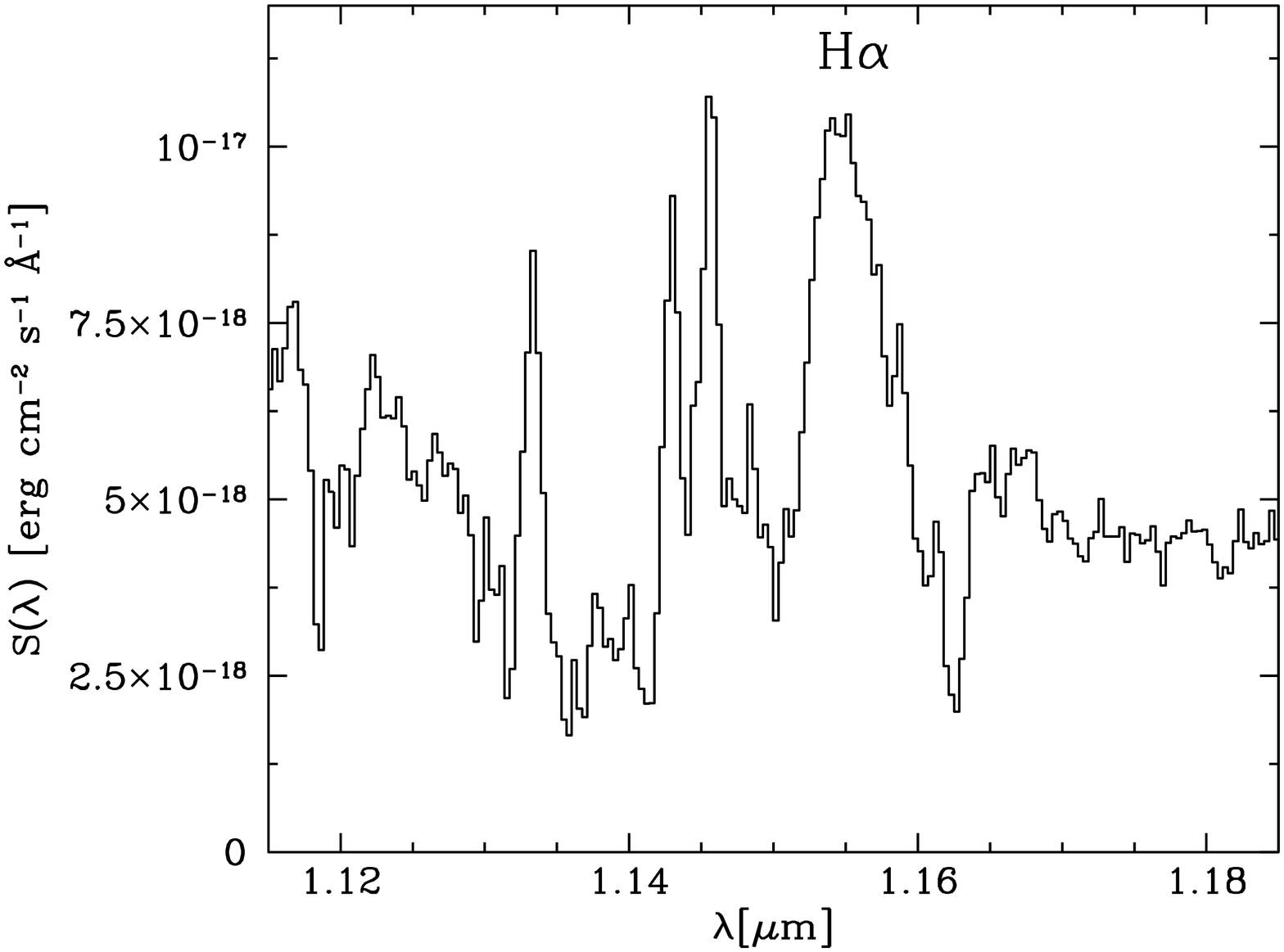}}}
\subfigure[HDF--S: s82.]{
\label{fig:spettro_s82J_isaac}
\rotatebox{-90}{
\includegraphics[height=0.48\textwidth]{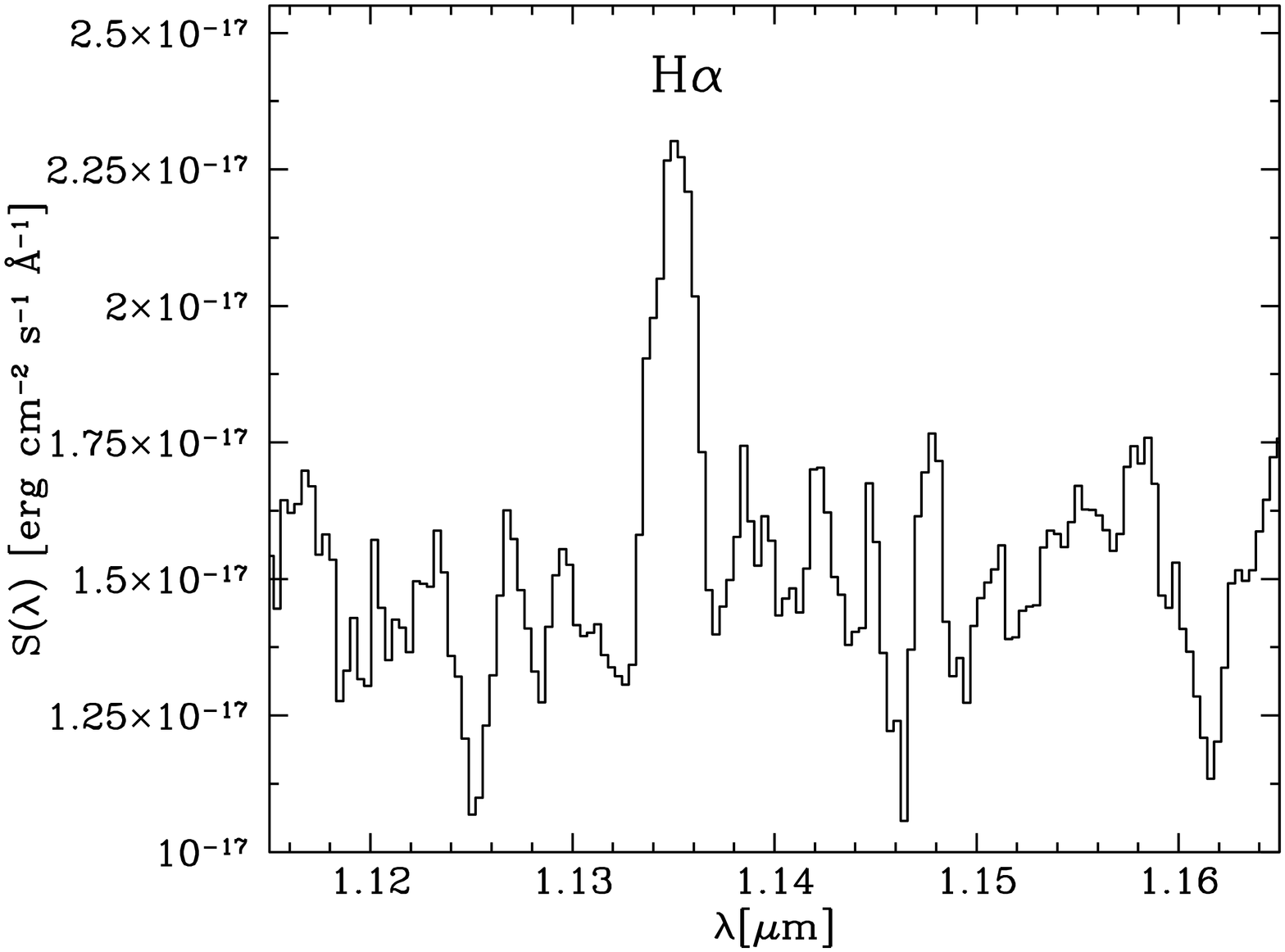}}}
\caption{Spectra of the 15 $\mu$m sources in the HDF--S observed with ISAAC at
the VLT.}
\label{fig:spettri_isaac}
\end{figure*}

\addtocounter{figure}{-1}
\setcounter{subfigure}{4}

\begin{figure*}[!ht]
\centering
\subfigure[HDF--S: s43.]{
\label{fig:spettro_s43J_isaac}
\rotatebox{-90}{
\includegraphics[height=0.48\textwidth]{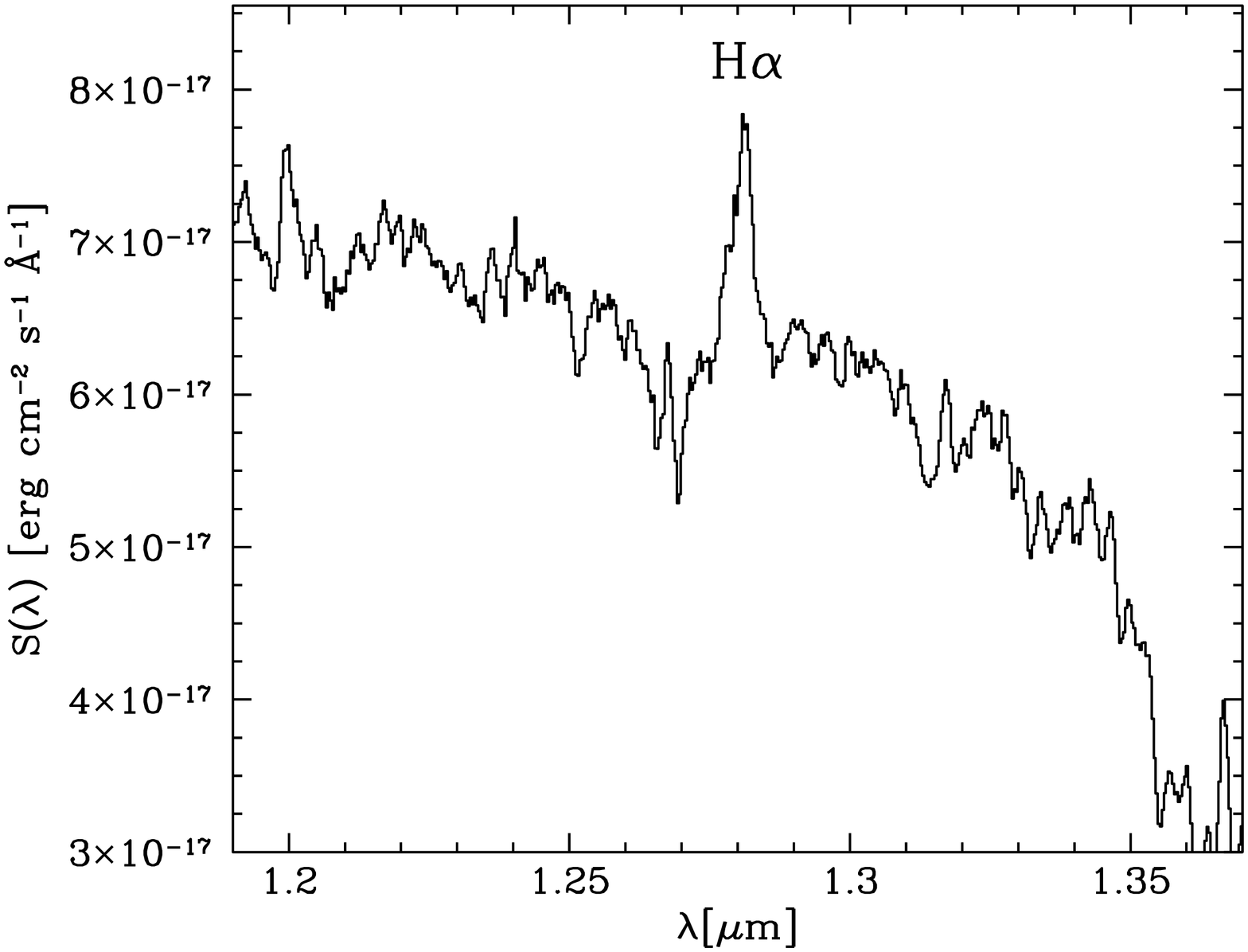}}}
\subfigure[HDF--S: s43.]{
\label{fig:spettro_s43H_isaac}
\rotatebox{-90}{
\includegraphics[height=0.48\textwidth]{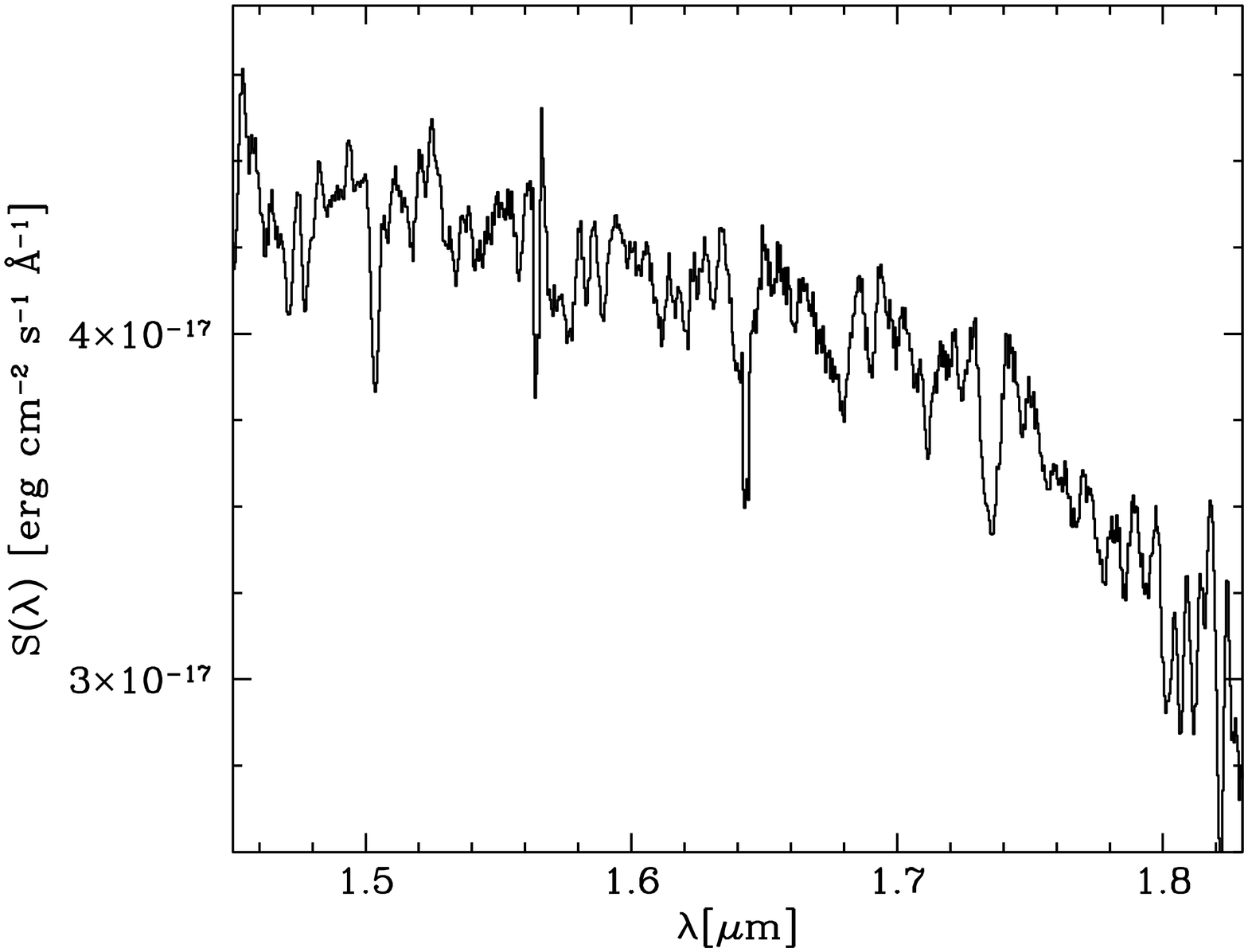}}}
\subfigure[HDF--S: s72.]{
\label{fig:spettro_s72SZ_isaac}
\rotatebox{-90}{
\includegraphics[height=0.48\textwidth]{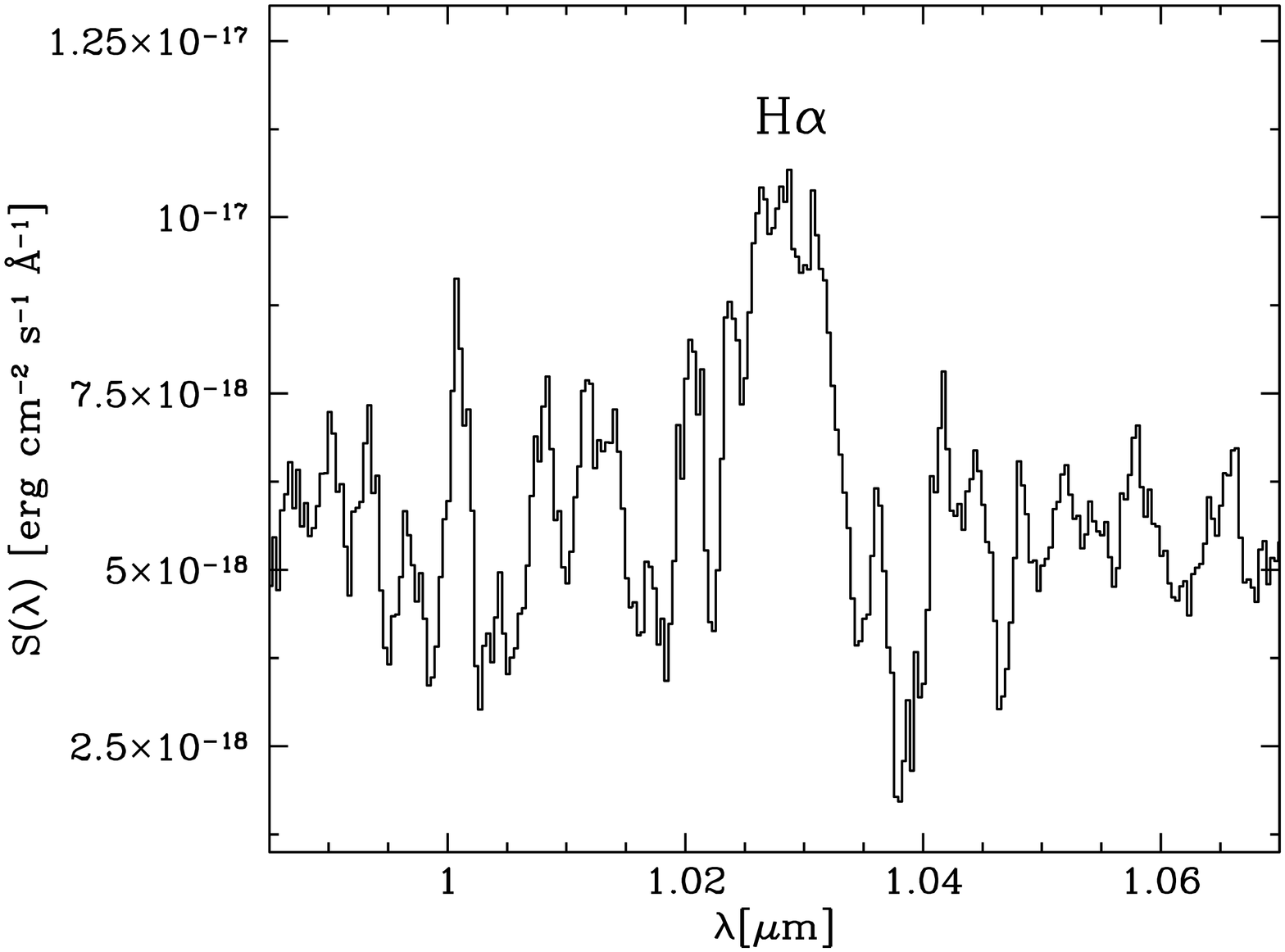}}}
\subfigure[HDF--S: s79.]{
\label{fig:spettro_s79J_isaac}
\rotatebox{-90}{
\includegraphics[height=0.48\textwidth]{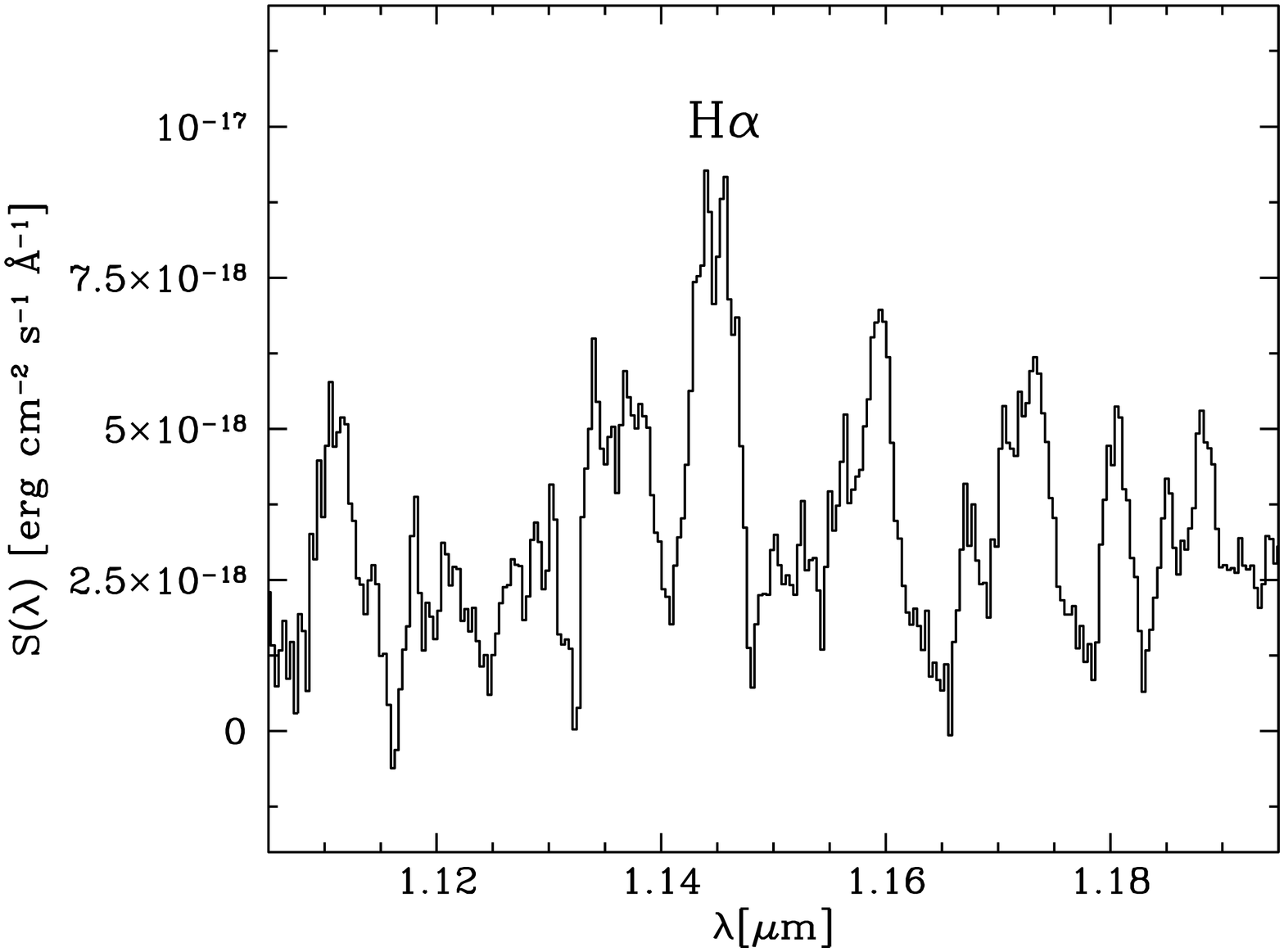}}}
\begin{flushleft}
{\bf Fig. \ref{fig:spettri_isaac}}: continues.
\end{flushleft}
\end{figure*}

\addtocounter{figure}{1}
\setcounter{subfigure}{0}

\begin{figure*}[!ht]
\centering
\subfigure[HDF--S: s14.]{
\label{fig:spettro_s14_emmi}
\rotatebox{-90}{
\includegraphics[height=0.48\textwidth]{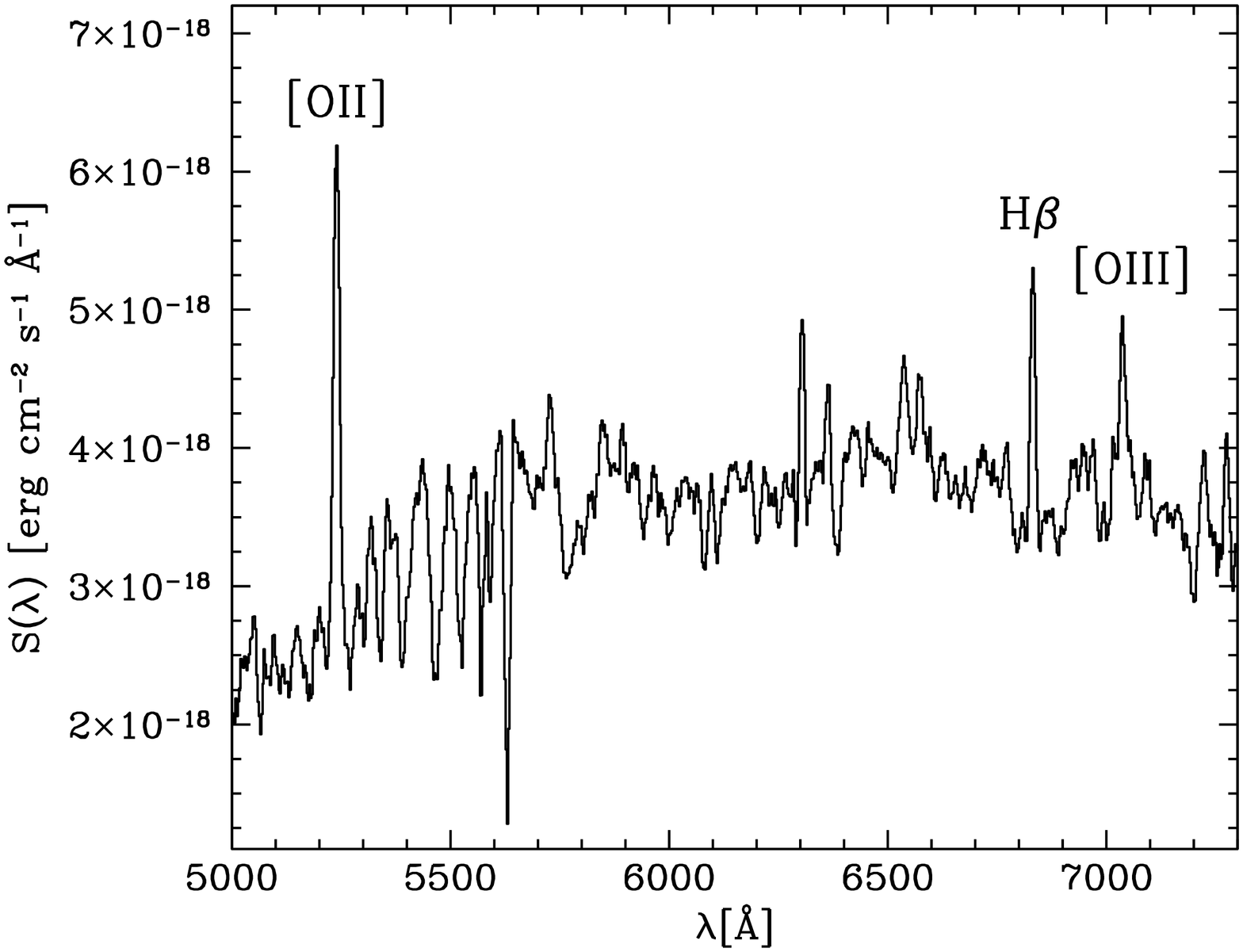}}}
\subfigure[HDF--S: s20.]{
\label{fig:spettro_s20_emmi}
\rotatebox{-90}{
\includegraphics[height=0.48\textwidth]{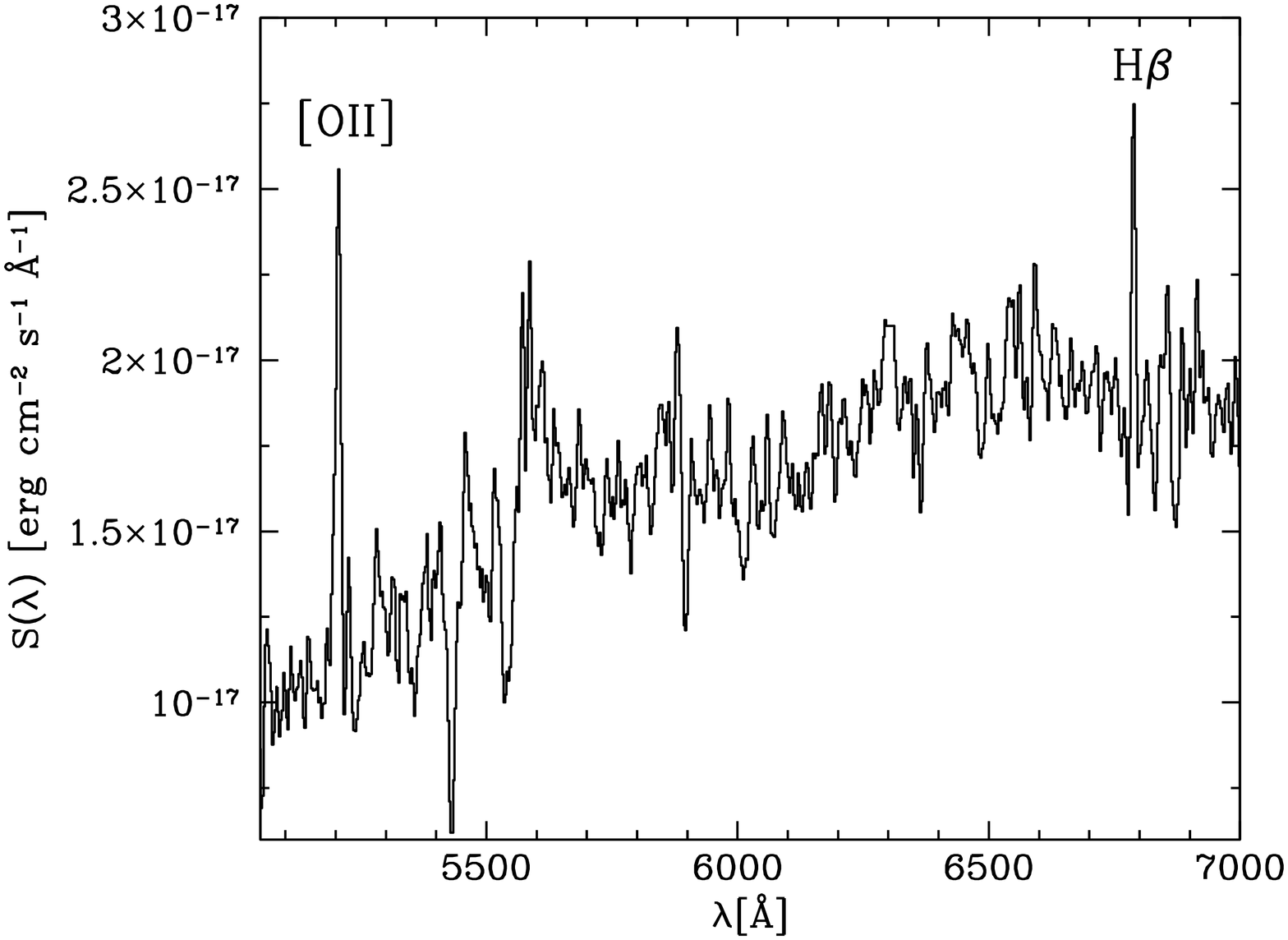}}}
\subfigure[HDF--S: s23.]{
\label{fig:spettro_s23_emmi_bis}
\rotatebox{-90}{
\includegraphics[height=0.48\textwidth]{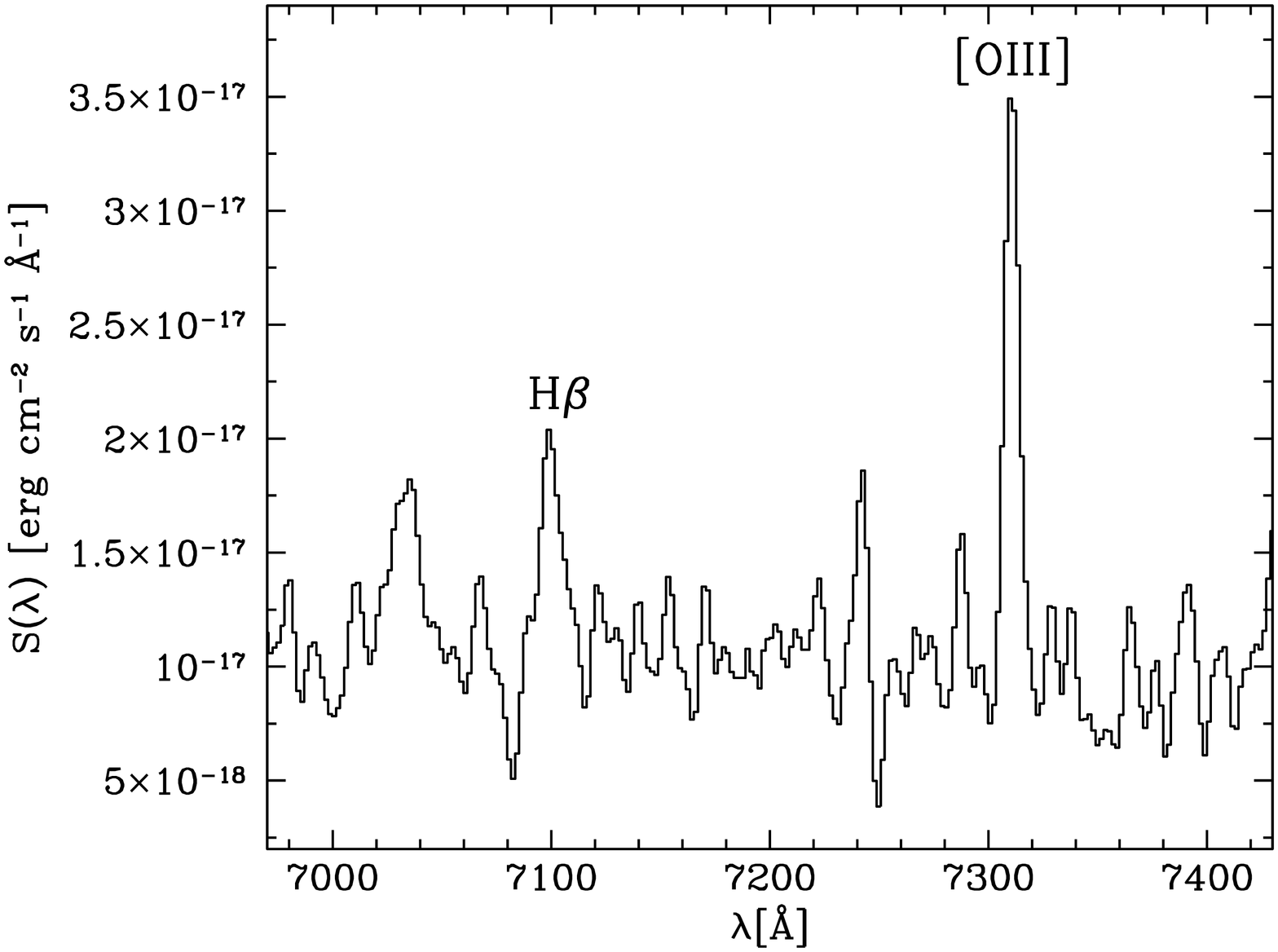}}}
\subfigure[HDF--S: s19, FORS1 spectrum.]{
\label{fig:spettro_s19H_fors1}
\rotatebox{-90}{
\includegraphics[height=0.48\textwidth]{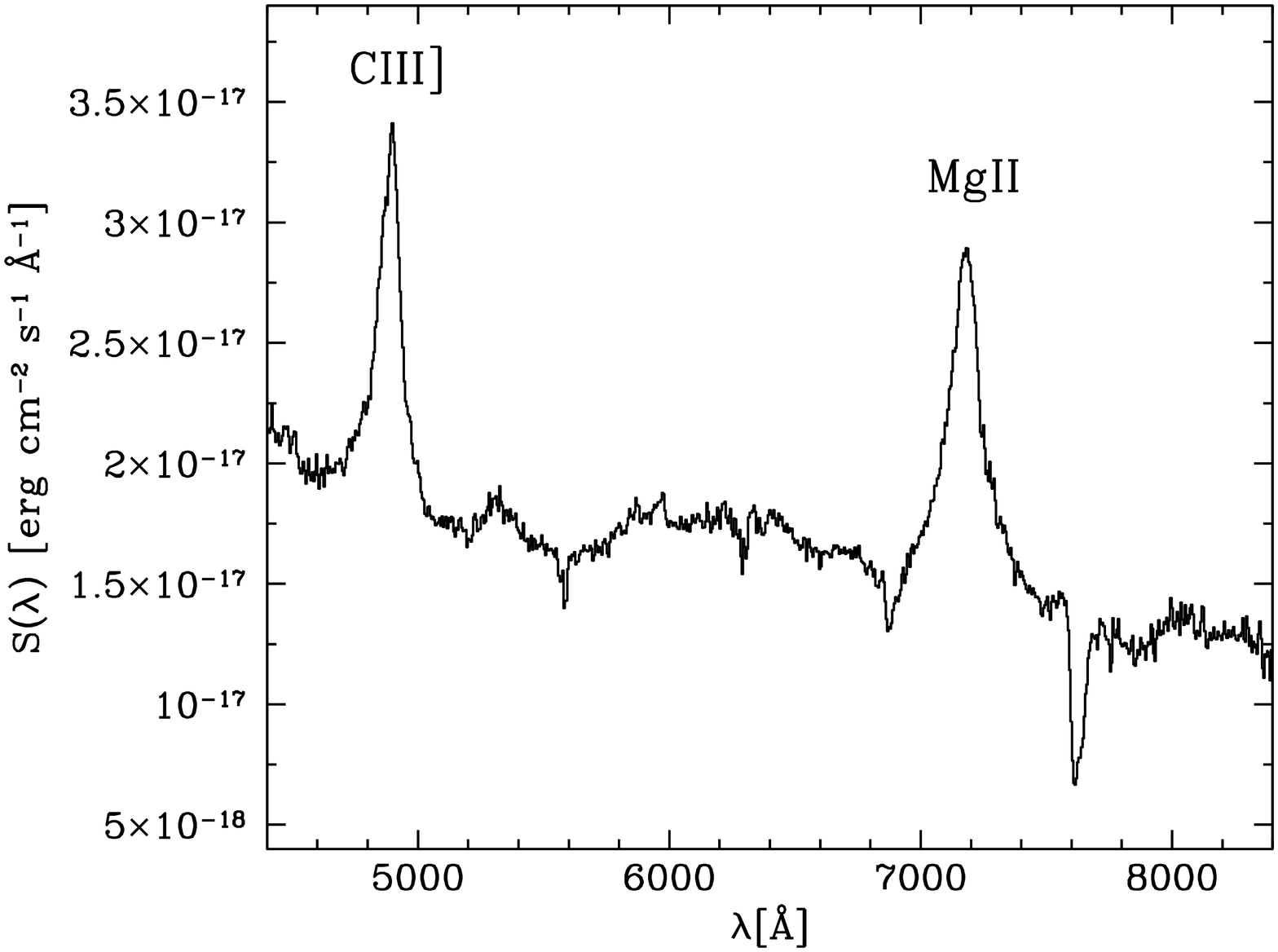}}}
\caption{Observed optical spectra: panels {\bf
(a)} to {\bf (c)} show 15 $\mu$m sources observed at NTT/EMMI. Panel {\bf (d)}
represents the QSO {\em s19} from FORS1/VLT: the two bright
broad lines are  Mg{\sc ii}$\lambda 2798$ and {\sc Ciii]}$\lambda 1908$,
redshifted at  $z\simeq1.56$ (courtesy of Rigopoulou et al., 2002, in
preparation).} 
\label{fig:spettri_emmi}
\end{figure*}

The data were reduced using applications in the ECLIPSE
(Devillard 1998) and IRAF\footnote{The package IRAF is distributed by the 
National Optical Astronomy Observatory which is operatedby the Association
of Universities for Research in Astronomy, Inc., under cooperative agreement
with the National Science Fundation.} packages.     Accurate sky subtraction is
critical to the detection of faint lines.  Sky was removed by
subtracting the pairs of offset frames. In some cases this left a
residual signal (due to temporal sky changes) which was then removed
by performing a polynomial interpolation along the slit.  OH sky
emission lines were also carefully removed from the spectra.  Spectrum
extraction for  each galaxy was performed using the APEXTRACT package.
Standard wavelength calibration was applied.

The ISAAC spectra have been flux-calibrated using standard infrared stars
from \cite{pickles1998} and \cite{vanderblik1996}.
The near-infrared spectra from the 1999 run have appeared in Rigopoulou et al.
(2000). Here we present in Figure \ref{fig:spettri_isaac} the ISAAC spectra 
from the 2000 and 2001 runs, while the measured fluxes are summarized in Table
\ref{tab:dati_spec}.

\subsection{Complementary optical spectroscopy}

To add further constraints on the nature of the 15 $\mu$m source population, 
particularly to estimate dust-extinction from Balmer line ratios, two 
different sets of ground--based optical spectroscopic observations have been analyzed.

\cite{tresse1999} observed the region around QSO J2233--606, which is in the
STIS HDF--South, with the ESO Multi Mode Instrument (EMMI, D'Odorico 1990) 
at NTT, during the
nights between September 23$^{th}$ and 25$^{th}$ and October 17$^{th}$ to 19$^{th}$
1998.  The data were obtained in Multi Object Spectroscopy (MOS) mode, with
different pointings and masks, for a total of 6 sets and roughly 200 slit
positions. 
\cite{tresse1999} chose slits with $1.02''$ or $1.34''$ width and spectral
resolutions of 10.6 and 13.9 \AA\, in the I band (centered at 7985\AA\,).

We also obtained observations with FORS2 (Appenzeller et al. 2000) at 
VLT/UT2 in MOS configuration, during the FORS2 commissioning phase, between December 
22, 1999, and January 5, 2000. The $300I$ grism
with a dispersion of $2.5$\AA$/pixel$ between 6000 and 11000 \AA\, and a $1''$-wide
slit were used. A few of the ISOCAM targets in our list have been included by chance 
in this spectroscopic survey.

Both EMMI and FORS2 data were reduced with the standard IRAF's tasks.
Pre--reduction (bias subtraction  and flat--fielding), extraction and wavelength 
calibration were performed with the usual procedure.
EMMI data were flux-calibrated using stars from \cite{stone1983} and
\cite{baldwin1984}, observed within each mask.  
Unfortunately, for FORS2 observations no data on standard stars were available.
However, two stars have been included in the MOS HDF--South frames:
after identification in the EIS catalogue, the spectral data were
calibrated by imposing to the observed stellar spectra to reproduce the I--band 
fluxes, after convolution with the instrument response function reported in 
the ESO exposure--time calculator.

Optical spectra with EMMI are reported in Figure \ref{fig:spettri_emmi}.

\section{SOURCE PROPERTIES BASED ON THE EMISSION LINES}
\label{par:spectra}

The ISAAC near--IR spectroscopy has been targeted to detect the redshifted
H$_\alpha$ line, while optical observations with EMMI and FORS2 
have allowed the study of the rest--frame emission lines up to $\sim 5000$\AA.

Table \ref{tab:dati_spec} summarizes the properties of reliable line 
detections with the three instruments. The columns report, in the order,
the source identification, the instrument, integration times, the
measured fluxes of {\sc [Oii]}$\lambda 3727$,
H$_\beta$, {\sc [Oiii]}$\lambda \lambda 4949, 5007$, H$_\alpha$+{\sc [Nii]}. 
The Table includes revised fluxes from the \cite{rigo2000} observations.
However, only the extremely good seeing conditions of the 1999 observing run have
allowed the separation of the [NII] from H$_\alpha$ for some of the sources. 
The corresponding separate fluxes for the resolved lines are reported in \cite{rigo2000}.
If an object was detected in different runs, both results are shown.
Uncertainties on the measured fluxes are of the order of 10\%.

\subsection{Aperture corrections}

We have applied aperture correction factors to account for the light missed
due to our finite 1'' size slit. We have calculated different
correction factors for unresolved (source sizes $\leq$ 1'')
and resolved sources (size $\geq$ 1''). 

For the unresolved sources we applied an average correction factor of 1.15 
which was calculated as follows.
We have smoothed the WFPC2 F814 image of the galaxies to a resolution of
0.6'' (to simulate the seeing during the observations) and measured
the ratio between the total source counts and the counts through an
artificial 1'' slit.

For the resolved sources (S27, S28, S53 ands S55) aperture corrections
have been calculated with two different methods. 

(1) We used the WFPC2 F814 image (smoothed as before to a resolution of 0.6'' to
account for the seeing) and carried out aperture photometry for
various aperture sizes (which allowed us to probe the distribution of
the flux and the source size in the sky).
We then calculated the flux through an
artificial 1'' slit oriented along the same PA as for the
observation. The average ratio between the total flux and that within the 1''
slit gave us an average correction factor of $\sim$1.7. 

(2) As a check, we have compared the broad band magnitudes in the z and J bands 
(corresponding to the H$\alpha$ rest-wavelength for z$\sim$0.8 and z$\sim$0.6, 
respectively) with the total observed H$\alpha$ line flux emitted by the sources.
Since we cannot estimate the line-to-continuum ratio from our spectra due to the 
low continuum signal, the ratio of the broad-band flux to the H$\alpha$ flux
provides a conservative upper limit to the aperture correction. 

The correction factors that we obtained in this way are $\sim$ 2 on average
for the four sources, consistent with the previously estimated average 
value.    In what follows we used correction factors of
1.15 for the unresolved and 1.7 for the resolved sources.

\subsection{Evidence for AGN contributions to the line emission}

A primary motivation for our spectroscopic follow up of the faint ISO sources 
was to determine to what extent the luminous IR emissions are contributed by
 AGNs, and how much by young stars. To this end, the observed optical
and IR spectra have been used to identify broad lines components and anomalous
line flux ratios.

Our search for broad permitted lines was limited by the low spectral resolution
of the ISAAC spectra, and even more by the low signal-to-noise ratio preventing reliable
setting of the underlying continuum.
The spectral resolving power of R$_{s}\simeq$600 corresponds to a rest-frame
resolution of 12$\AA$ at z=0.6 and a velocity 
threshold of $\sim 250$ km/s for the ISAAC Low-Resolution gratings. For the
Medium-Resolution grating the resolving power corresponds to $\sim 40$ km/s.
Consequently, for 16 out of 20 ISO sources in our sample observed at low-resolution,
only broad-line components (v$>1000$ km/s) can be resolved in principle.
For S14, S20 and S23 we have good quality optical NTT/EMMI spectra with resolving 
power $R_s=580$ ($\sim500\,km/s$).

We report in Table \ref{widths} the half-power widths of the relevant permitted
lines based on both IR and optical spectra. Only for one source (S19) 
these imply velocity fields significantly in excess of 1000 km/s (indicative of the 
presence of an AGN). For four more sources with H$_\alpha$+{\sc [Nii]} line
widths also formally in excess of 1000 km/s, the poor S/N and the unresolved 
{\sc [Nii]} contribution prevent a reliable measure of the velocity field.
In all other cases the line widths are consistent with those typical of massive 
starburst galaxies.

\begin{table}[!ht]
\centering
\begin{tabular}{c c c c c}
\hline
\hline
Obj. & line & FWHM & FWHM & Threshold  \\
id. &  meas.& [\AA]& [km/s] & velocity \\
\hline
s14 & H$_\beta$ & 11 & 481   & 300  \\
s19 & H$_\alpha$& 261 & 4642 & 400 \\
s20 & H$_\beta$ & 10 & 443   & 300  \\
s23 & H$_\beta$ & 14 & 591   & 400  \\
s25 & H$_\beta$ & 10 & 390   & 200  \\
s27 & H$_\alpha$ & 8 & 231   & 300  \\
s38 & H$_\alpha$ & 32 & 612  & 250 \\
s39 & H$_\alpha$ & 70 & 1200  & 450 \\
s40 & H$_\alpha$+{\sc [Nii]} & 25 & 503  & 350 \\
s43 & H$_\alpha$+{\sc [Nii]} & 50 & $<$1172 & - \\
s55 & H$_\alpha$+{\sc [Nii]} & 50 & $<$1299 & - \\
s60 & H$_\alpha$+{\sc [Nii]} & 18 & 368  & 300   \\
s62 & H$_\alpha$+{\sc [Nii]} & 19 & 502  & 500   \\
s72 & H$_\alpha$+{\sc [Nii]} & 84 & $<$2477 & - \\
s79 & H$_\alpha$+{\sc [Nii]} & 27 & 709    & 500 \\
s82 & H$_\alpha$+{\sc [Nii]} & 26 & 703    & 500 \\
\hline
\end{tabular}
\caption{Widths of relevant emission lines from ISAAC Low-Res and optical spectroscopy.
The source name, reference line used for the measure, FWHM in \AA, FWHM in km/s, and
the threshold resolving power in km/s are reported.}
\label{widths}
\end{table}

The H$_\alpha$ to {\sc [Nii]} flux ratio may be used in principle as an indicator 
of the ionization field in the source. 
However, because of our limited resolving power, the line complex was resolved for 
only 3 sources of the 1999 run (Rigopoulou et al. 2000). Interestingly, one of these
(S38 at z=1.39)
reveals an inverted line ratio (H$_\alpha$/{\sc [Nii]}$\simeq$0.8), together with a 
moderately broadened H$_\alpha$ ($\sim 600$ km/s, see Table \ref{widths}). 

No evidence for broad components or peculiar H$_\alpha$/{\sc [Nii]} flux ratios 
was found in the optical counterparts to S25, S27 and S55 observed in the ISAAC 
Medium-Resolution mode, whose spectra are consistent with those of standard spiral 
and starburst galaxies with no AGN signatures (Rigopoulou et al. 2002).

Altogether, two out of 21 of the faint ISO sources in our spectroscopic survey 
reveal evidence for either type-I (source S19) or type-II (source S38) 
AGN activity.
These indications from line measurements will be compared in Sect.\ref{par:AGNc}
with those coming from the study of the optical-IR-radio continuum SEDs.


\begin{table}[!ht]
\footnotesize
\begin{tabular}{ l c r@{.}l  c c r@{.}l }
\hline
\hline
{\em Obj.} & {\em $L^\ast$} &  \multicolumn{2}{c}{\em $SFR$} & {\em $A_V$} &
{\em $A_V$}  &  \multicolumn{2}{c}{\em $SFR$} \\ 
\#  & (H$\alpha$) & \multicolumn{2}{c}{(H$\alpha$)} &  (col.) &  (spec.) & \multicolumn{2}{c}{(H$\alpha$), corr}  \\ 
\hline
 16$^\dagger$	&	0.251 &       1&983  &1.00    & --    & 3&935		     \\
 23$^\dagger$	&	$>$0.194 &       $>$1&532  &2.40    & 1.84  & $>$5&405		     \\
 25$^\dagger$	&	0.569 &       4&495  &--      & 2.10  & 18&964		     \\
 27$^\dagger$	&	0.867 &       6&856  &1.98    & 1.69  & 21&874		     \\
 28$^\dagger$	&	0.191 &       1&504  &1.20    & --    & 3&423		     \\
 38$^\dagger$	&	3.060 &      24&185  &2.60    & --    & 143&865		     \\
 39$^\dagger$	&	8.930 &      70&541  &2.45    & --    & 378&465		     \\
 40		&	1.678 &      14&479  &--      & --    & \multicolumn{2}{c}{--}   \\
 43		&	2.358 &      18&630  &2.36    & --    &   93&955  		  \\
 53$^\dagger$	&	1.622 &      12&819  &1.42    & 2.05  &  52&443  		  \\
 55$^\dagger$	&	1.234 &       9&751  &1.83    & 2.63  &  59&125  		  \\
 60$^\dagger$	&	3.158 &      24&944  &2.25    & --    &  116&725  		  \\
 62$^\dagger$	&	0.812 &       6&414  &1.40    & --    &  16&756  		  \\
 72		&	0.858 &       6&782  &--      & --    & \multicolumn{2}{c}{--}   \\
 73		&	0.207$^\times$&1&635 &--      & --    & \multicolumn{2}{c}{--}   \\
 79		&	0.436 &       3&449  &--      & --    & \multicolumn{2}{c}{--}   \\
 82		&	0.984 &       7&773  &--      & --    & \multicolumn{2}{c}{--}   \\
\hline
\multicolumn{8}{l}{$^\dagger$: H$\alpha$ fluxes from \cite{rigo2000}}\\
\multicolumn{8}{l}{$^\ast$: H$\alpha$ luminosities in units of $10^{42}$ erg s$^{-1}$}\\
\multicolumn{8}{l}{$^\times$: S73, no aperture correction performed.}
\end{tabular}
\caption{Relevant spectroscopic data: aperture-corrected H$\alpha$ luminosities, 
rates of star formation based on H$\alpha$, and extinction estimates from the Balmer
decrement. In the fourth column we report extinction values based on photometric fits
with spectral synthesis codes (see Sect. 4.1) and values of SFR corrected for 
slit-aperture and reddening (from Balmer lines when available).  
SFRs are in $M_\odot$ yr$^{-1}$.} 
\label{tab:sfr_extinction}
\normalsize
\end{table}

\subsection{Estimates of dust extinction from line ratios}

Corrections of the H$_\alpha$ flux for dust extintion are needed
when using the line fluxes to estimate the rates of star formation.
These are usually computed from the observed ratios of the Balmer lines compared
with theoretical models of atomic transitions and nebular emission.
\cite{hummer1987} report the ratios of emission lines based on Case B recombination 
theory for $T=10000$:
$$
\frac{I(\textrm{H}\alpha)}{I(\textrm{H}\beta)}=2.85 .
$$
By comparing this value with the observed line ratios and using a standard 
extinction law \citep{fitzpatrick1999}, we obtain
the $A_V$ values reported in Table \ref{tab:sfr_extinction}.
The same Table also includes the extinction values based on fits of spectrophotometric
models to the optical/near-IR SEDs, as explained in Sect. \ref{par:Av_cont}.

All $A_V$ values turn out to be greater than $1.5$ magnitudes, implying 
substantial amounts of dust extinction in these objects, which are not typical of 
normal quiescent spirals for which $A_V\simeq 0.3-0.5$.
If anything, these estimates are likely to be lower limits, since this indicator 
based on Balmer line ratios quickly saturates in a mix of dust and radiative
sources.

\subsection{Estimates of the Star Formation Rate}

Nebular emission lines, such as H$_\alpha$, are generated in the interstellar
medium, ionized by the ultraviolet continuum of the young stars recently formed
in the starburst (those with ages $\leq 10^7$ yrs). 
These lines then provide a direct indication of the rate of the
ongoing stellar formation. 


\begin{table}[!ht]
\centering
\scriptsize
\begin{tabular}{ l c l l l c}%
\hline
\hline
{\it Obj} & {\it z} & $L_{IR}$  & {\em SFR}            &  {\em M}    & {\em M range}       \\
 {\em \#} &         & $L_\odot$ & $[M_\odot\,yr^{-1}]$ &$[10^{11}M_\odot]$ &$[10^{11}M_\odot]$ \\
\hline
14 & 0.41 & $5.8\ 10^{10}$       & 10.0 & 0.32       & 0.20$\div$0.40 \\
15 & (0.55)&$1.8\ 10^{11}$       & 30.6 & 3.10       & 2.30$\div$3.10 \\
16 & 0.62 & $8.7\ 10^{10}$       & 14.9 & 0.29       & 0.21$\div$0.33 \\
18 & (0.55)&$1.4\ 10^{11}$       & 23.4 & 5.20       & 5.00$\div$5.50 \\
19 & 1.57 & $1.4\ 10^{12\ast}$   &  --  &  --        & -- \\
20 & 0.39 & $3.5\ 10^{10}$       & 6.0  & 0.80       & 0.55$\div$1.28 \\
23 & 0.46 & $2.9\ 10^{11}$       & 50.2 & 0.97       & 0.42$\div$1.50 \\
25 & 0.58 & $3.3\ 10^{11}$       & 56.0 & 0.80       & 0.50$\div$1.20 \\
27 & 0.58 & $2.6\ 10^{11}$       & 44.8 & 4.70       & 3.90$\div$5.70 \\
28 & 0.58 & $9.6\ 10^{10}$       & 16.5 & 0.40       & 0.25$\div$0.70 \\
30 & (0.40)&$2.7\ 10^{10}$       & 4.6  & 0.01       & 0.005$\div$0.020\\
36 & (0.65)&$2.7\ 10^{11}$       & 46.3 & 0.40       & 0.20$\div$0.60 \\
38 & 1.39 & $1.3\ 10^{12\ast}$   &  --  & 1.40	     & 1.00$\div$3.00 \\
39 & 1.27 & $4.4\ 10^{12}$       & 748.9& 1.70       & 1.00$\div$3.20 \\
40 & 1.27 & $1.5\ 10^{12}$       & 264.9& 1.20       & 0.70$\div$3.00 \\
41 & (0.30)&$1.6\ 10^{10}$       & 2.7  & 0.055      & 0.045$\div$0.069\\
43 & 0.95 & $3.3\ 10^{11}$       & 57.5 & 0.50       & 0.20$\div$1.00 \\
45 & (0.65)&$3.9\ 10^{11}$       & 66.6 & 0.80       & 0.40$\div$1.60 \\
48 & (0.30)&$1.6\ 10^{10}$       & 2.8  & 0.10       & 0.06$\div$0.20 \\
52 & (0.60)&$1.2\ 10^{11}$       & 20.1 & 0.13       & 0.11$\div$0.16 \\
53 & 0.58 & $2.2\ 10^{11}$       & 38.5 & 1.20       & 1.00$\div$1.50 \\
55 & 0.76 & $2.6\ 10^{11}$       & 45.4 & 1.30       & 0.60$\div$1.60 \\
60 & 1.23 & $9.7\ 10^{11}$       & 167.2& 2.50       & 1.60$\div$4.00 \\
62 & 0.73 & $2.2\ 10^{11}$       & 36.9 & 0.66       & 0.35$\div$0.90 \\
67 & (1.00)&$1.4\ 10^{12}$       & 244.3& 0.34       & 0.21$\div$0.60 \\
71 & (0.45)&$2.6\ 10^{10}$       & 4.4  & 0.04       & 0.015$\div$0.060\\
72 & 0.55 & $2.2\ 10^{11}$       &  36.9& 1.40       & 1.10$\div$1.80 \\
73 & 0.17 & $8.0\ 10^{10}$       & 13.7 & 1.50       & 0.90$\div$1.90 \\
75 & (0.45)&$1.5\ 10^{11}$       & 25.0 & 1.30       & 0.93$\div$1.60 \\
77 & (0.40)&$6.9\ 10^{10}$       & 11.8 & 0.80       & 0.48$\div$1.15 \\
79 & 0.74 & $2.2\ 10^{11}$       & 38.4 & 0.60       & 0.30$\div$1.20 \\
82 & 0.69 & $5.2\ 10^{11}$       & 90.2 & 0.50       & 0.22$\div$1.10 \\
85 & (0.40)&$2.8\ 10^{10}$       & 4.7  & 5.70       & 5.20$\div$6.00 \\
\hline
\multicolumn{6}{l}{
$^\ast$ Based on the AGN model described in Appendix A.}
\end{tabular}
\footnotesize
\caption{Values of the baryonic masses, SFRs and bolometric IR (8$\div$1000
$\mu$m) luminosities of the 15 $\mu$m sources in the HDF--S, 
as derived from the analysis of their spectral energy distributions 
(see Sects. 4.3 and 4.4). 
}
\label{tab:sfr_masses_ir}
\end{table}

The conversion factor between ionizing flux and SFR may be computed with an 
evolutionary synthesis model, assuming solar abundances, a Salpeter IMF 
(0.1--100 M$_{\odot}$) and for continuous bursts with duration in the range between 
few$\times$10$^{7}$ yrs to few$\times$10$^{8}$ yrs. 
We adopt for this the Kennicutt (1998) relation:  
\begin{equation}
\frac{SFR}{M_\odot yr^{-1}}=7.9 \cdot10^{-42}L_{\textrm{H}\alpha}(\textrm{erg
s}^{-1}),
\label{SFR}
\end{equation}
which we used to derive the values of ongoing SFR in the HDF--S objects with
reliable H$_\alpha$ detections.
After conversion from fluxes to luminosities and after applying slit-aperture and 
extinction corrections, SFRs were calculated from eq. (\ref{SFR}) and reported in Table
\ref{tab:sfr_extinction}.

\section{SOURCE PROPERTIES BASED ON THE CONTINUUM EMISSION}\label{par:continuum}

The extensive spectral coverage of our source SEDs between the UV and the far-infrared
(including the radio flux for a few) allows adding very significant
constraints on the physical processes taking place inside them.

For typical IR galaxies the bulk of stellar formation happens inside dust-rich and
optically thick molecular clouds, absorbing the UV--optical continuum emitted 
by luminous young stars and re-emitting it in the infrared. 
Given the large extinction values, the optical to near-IR continuum spectrum 
is moderately influenced by the starburst and includes important contributions by the 
less-extinguished older stellar populations (e.g. Poggianti, Bressan, Franceschini 2001). 
On the contrary, the mid- and far-IR spectrum is typically
dominated by thermal emission by dust mostly illuminated by the young stars
(the radio flux is also proportional to the number of recently born massive stars).

As a consequence, our capability to cover simultaneously the whole SEDs up to the far-IR
provides a unique opportunity to sample all stellar generations and ages in these
galaxies, and in particular to compare the mass fractions in old stars (the inactive, 
passively evolving component) to those of newly formed stars (the "active" component),
hence to evaluate the "activity" level in these galaxies.

SEDs and (when available) WFPC-2 images of our sample sources are reported in Figs. 
A.2 and A.3, together with spectral fits based on GRASIL, 
and with the SEDs of template galaxies used for comparison.
The dot-dashed line in each panels, in particular, corresponds to the SED of 
M51, which we assumed to represent the prototype inactive spiral: the comparison of this
template, scaled to fit the near-IR spectrum, with the ISO data at 6.7 and 15 $\mu$m 
shows that quite often the observed mid-IR fluxes are in excess by a substantial factor. 
Assuming that this excess IR emission is due to the starburst, this factor
may be taken as a measure of the level of "activity" in the ISO galaxies.

Additional contributions to the source fluxes - further enhancing this
"activity" level - may come from nuclear non-thermal emission of gravitational 
origin, an AGN component.

\subsection{Estimates of dust extinction based on the continuum}\label{par:Av_cont}

The lack of information on the Balmer line ratios for several of our sample sources
has forced us to look for alternative methods to estimate the dust extinction.
A widely used one is to exploit colour indices. Using the codes of spectrophotometric
synthesis GRASIL (Silva et al., 1998) and STARS (Sternberg 1998, Thornley et al. 2000)
we have generated a grid of models for various star formation histories and
bursts of different duration, and calculated the intrinsic (dust-free) V--K colours 
(typically varying in the range V--K = 1.1--1.5). 
We have then applied infrared and optical
K-corrections from Poggianti (1997) and Coleman (1980),
respectively. By comparing the observed and model predicted V--K
colours, we obtained a median A$_{V}$ of 1.8, assuming a screen model
for the extinction. We stress that the extinction estimates based on
the V--K colour represent a global (galaxy-wide) extinction which may
not necessarily represent the extinction towards the deeply embedded
young stars where most of the line emission originates. 
Note that, although somewhat more physically motivated, a model with homogeneous mix
of dust and stars could not provide a fit to the optical spectrum: at increasing
the intrinsic dust optical depth the extinction saturates to $A_V\simeq 0.7$ (e.g. Poggianti, 
Bressan, Franceschini 2001), a value not large enough to explain our typical source 
spectra.

Alternatively, we can estimate the extinction based on the rest-frame
L$_{\small \rm UV}$(2800) continuum. Values of L$_{\small \rm UV}$ have been
interpolated from B V I colours, using grids of synthetic models as described above,
with constant star formation rate.  The median L$_{\small \rm UV}$(2800) for the
present ISOHDF-S sample is 1.47$\times$10$^{40}$ erg s$^{-1}$ A$^{-1}$. 
For the same IMF as assumed before, we find the following relation between the SFR
and L$_{\small \rm UV}$ (2800):
\begin{equation}
SFR(M_{\odot}/yr) = 1.88 \times10^{-40}L_{\small \rm UV} (2800)
(erg s^{-1} A^{-1}).
\end{equation}
The ratio between the H$_{\alpha}$ and UV-based SFRs turns out to be
SFR(H$_{\alpha})/$SFR(2800)$\sim$4.0. 
For the corresponding extinction it follows that
A(H$_{\alpha}$) = 0.76 A$_{\small V}$ and A(UV)=1.6 A$_{\small V}$
(Pei 1992).  We then deduce that, for a screen distribution, the extinction
is A$_{\small V}\sim$1.6. This extinction corresponds to a correction
factor for the SFR(H$_{\alpha}$) of $\sim$3.

\subsection{Disentangling AGN from starburst signatures in the UV-optical-IR-radio 
SEDs}\label{par:AGNc}

The mid-IR spectra of galaxies are a complex mixture of various components. 
Typically the most important one is the emission by molecular complexes 
(probably Polycyclic Aromatic Hydrocarbons, PAH, see Puget \& Leger 1989, but their 
nature is still uncertain) producing bands at 6.2, 7.7, 8.6 and 11.3 $\mu$m
on top of a hot dust continuum. The PAH emission is particularly 
prominent in star-forming galaxies. Both the continuum intensity and 
the PAH's equivalent widths are sensitive to the presence of an AGN: 
while the AGN luminous point-like source enhances the hot-dust continuum emission,
the PAH molecules tend to be destroyed by the intense radiative field and the 
corresponding emission bands become weaker (Lutz et al. 1998, Rigopoulou et al. 
1999, Tran et al. 2001). 

Therefore starburst-dominated and AGN-dominated galaxies tend to display different 
mid-IR spectra. The former show a bumpy spectrum with prominent emission features 
at $\lambda>6\ \mu$m and a steeply decresing flux shortwards of 6 $\mu$m, due to 
the moderate intensity of the radiative field and to the lack of very hot dust. 
A typical starburst spectrum
is reported in Figs. A.2 and A.3 as the thick continuous
line, corresponding to the SED of the galaxy M82.

On the contrary, AGN-dominated sources show rather flat mid-IR spectra, with almost
absent PAH features and strong emission by very hot dust detectable down to few
microns (a comparison of the spectrum of the Seyfert galaxy NGC 1068 with
those of starbursts is reported in Aussel et al. 1999 and Elbaz et al. 2002).

The shape of the mid-IR spectrum can then in principle be used to disentangle 
between the two power mechanisms. In our case we can exploit the 
ratio of the LW3 to LW2 fluxes as a measure of how fast the rest-frame SED drops  
at $\lambda<6\ \mu$m for sources at $z>0.4$: while this flux ratio is expected to
be $\geq 4$ for starbursts as a consequence of the LW2 flux missing the redshifted
PAH and dust emissions, the ratio becomes $\frac{S(LW3)}{S(LW2)}\leq 4$ 
in the case of an AGN-dominated source. 
Two such sources in Fig. A.2 are S19 and S38, having
respectively $\frac{S(LW3)}{S(LW2)}\simeq 1.5$ and 2.3.  For a third source, 
S82, the flux ratio is also low, $\frac{S(LW3)}{S(LW2)}\simeq 2.6$,
and this could also be a type-II quasar.
In all other cases, either LW2 has no detection or the flux ratio falls in the 
starburst regime.

Note that this starburst/AGN discriminant would not work for sources at $z<0.4$: 
an example is S73, showing a low $\frac{S(LW3)}{S(LW2)}\simeq 2$ value only
because its low redshift (z=0.17) prevents the PAH bundle to be redshifted
out of the LW2 band. 

Radio data could in principle add valuable information to the SED analysis. 
On one side, given the tight radio/far-IR correlation for star-forming galaxies 
(Helou, Soifer and Rowan-Robinson 1985), 
the radio flux may provide an independent estimate of the bolometric luminosity and
star formation rate. On the other side, a substantial excess of radio emission above
the value pertaining to the radio/far-IR relation may be taken as evidence for 
emission by an AGN.

Unfortunately, for only a handful of the HDF South sources data at 1.4, 2.5, 
4.9 and 8.6 GHz are reported by Mann et al. (2002) based on ATCA observations.
Of the four radio-detected sources in our sample, S19 was already identified as
a bona-fide type-I quasar, while S39 was suspected to include an AGN contribution
based on the broadness of the H$_\alpha$ line and morphology (see the Appendix). 
For S39 the 1.4 GHz radio emission is however consistent with the radio/far-IR 
relation for starbursts (see Fig. \ref{fig:sfr_radio_ir} below).

\begin{figure}[!ht]
\centering
\includegraphics[height=0.45\textwidth]{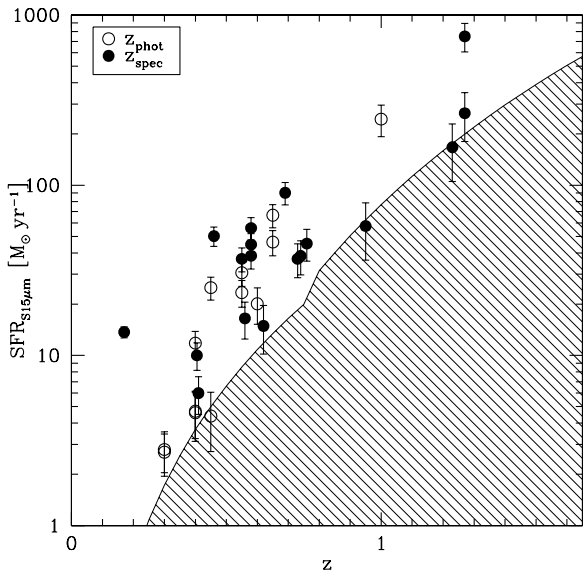}
\caption{Rates of star formation in the 15 $\mu$m sources based on 
the bolometric IR luminosity, as a function of redshift.
Filled and open circles refer to objects with spectroscopic or
photometric redshifts. 
The shaded area represents the unobservable region given by 
our sensitivity limit of 95 $\mu$Jy imposed by the sample selection. 
}
\label{fig:sfr_z}
\end{figure}

\begin{figure*}[!ht]
\centering
\includegraphics[width=.9\textwidth,height=0.7\textwidth]{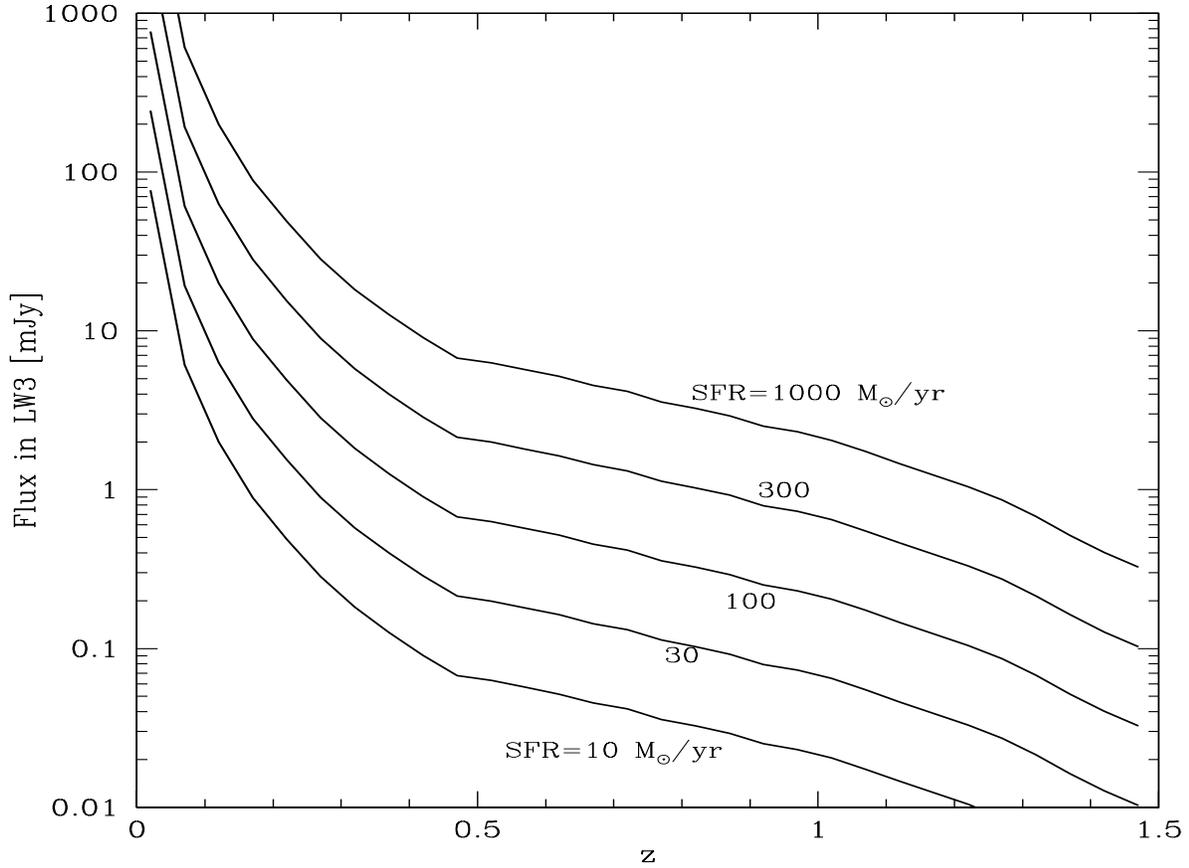}
\caption{Relation of the observed flux in the 15 $\mu m$ LW3 ISO band versus redshift,
as a function of the rate of star formation (SFR). This is based on the assumption
that the IR flux is dominated by dust-reprocessed emission of young stars.
The calibration of this relation has been obtained using the M82 spectral 
template, the detailed response function of the LW3 filter (see eqs. 5 and 7 in
Franceschini et al. 2001) and the Kennicutt's (1998) relation between SFR and 
L$_{IR}$ for a Salpeter IMF between 0.1 and 100 M$_\odot$.
}
\label{five}
\end{figure*}

For the two other radio detections, S23 and S73, the radio and mid-IR fluxes
are entirely consistent with a pure starburst emission (see again Tables
\ref{tab:sfr_masses_ir} and \ref{tab_sfr_radio}, and also Mann et al. 2002).

In summary, our analysis has found clear evidence for the presence of AGNs 
in two of the sample sources (the type-I S19, and the type-II S38). 
For a third source, S39, we suspect the presence of an AGN contribution 
because of morphology and a marginally broadened H$_\alpha$ line profile. 
For a fourth source, S82, the low LW3/LW2 flux ratio could indicate the
contribution to the mid-IR spectrum of a type-II AGN.
For all other HDF South sources any AGN contributions should be minor. 
Although the statistics is poor, this 10$\div$20\% AGN fraction among the 
faint ISO sources is the same as found by Fadda et al. (2002) using the hard 
X-ray emission as AGN diagnostic. This is also in agreement with the results by
Alexander et al. (2002) exploiting the 1 Msec Chandra X-ray exposure on the HDFN, 
in which 20 of the 41 ISOCAM sources were detected, 4 of which were classified
as AGNs and 15 as emission line galaxies.

\subsection{Estimating the Rate of ongoing Star Formation}\label{par:sfr}

Once the far-infrared activity of the source is reliably attributed to young 
stars, a fundamental indicator of the source physical status is 
the rate of ongoing star formation (SFR). 
Although UV-optical line and continuum emissions are contributed by newly born 
stars, Sanders \& Mirabel (1996) have shown that starbursts with bolometric 
luminosities above $10^{10} L_\odot$ produce the bulk of their energy in the far-IR. 
For this reason, the bolometric IR luminosity of a starburst galaxy is the most 
direct and reliable estimator of star formation. It is then obviously important for us 
to compare our previous estimates of the SFR based on the H$_\alpha$ flux
with more robust ones based on far-IR luminosities.

To this end, a measure of the source flux around 100 $\mu$m, where the galaxy IR 
SEDs are expected to peak, would be needed in principle. Unfortunately, this is not
currently available to us in the HDF South. The imaging capabilities of ISO are 
strongly limited at such long wavelengths by the poor spatial resolution and 
high confusion noise (e.g. Franceschini et al. 2001). There is neither much 
perspective of an improvement until the operation of FIRST-Herschel in 2007.
An important result by Elbaz et al. (2002), however, was to show 
that the mid-IR flux for a large variety of galaxies (from normal galaxies
to luminous and ultra-luminous dusty starbursts) 
is extremely well correlated with the bolometric IR emission.
They have shown that in a large sample of local objects only a very small fraction
($\sim$ few \%)
show significant departures from this correlation (a remarkable such discrepant case 
is Arp 220, an ultra-luminous IR galaxy [ULIRG] showing a shortage of mid-IR
flux compared with the far-IR one, due to dust self-absorption as evident from
the PAH spectrum [Rigopoulou et al. 1999, Haas et al. 2001]).

This almost linear correlation of the bolometric far-IR and mid-IR luminosities
was proven to hold in local galaxies by comparing the bolometric fluxes from IRAS data
with mid-IR fluxes in various channels, including the IRAS 12 $\mu$m and the ISO LW3 and 
LW2 bands: considering that for sources at $z\sim 1$ the rest-frame emission in LW2
shifts to the observed LW3 band, this analysis by Elbaz et al. 
proved that the correlation is likely to hold for galaxies at least up to $z\sim 1$.
This is also confirmed by the excellent match between the SFR estimates based 
on the mid-IR and radio fluxes for dusty starbursts at $z\leq 1$
(see Sect. \ref{5.1} below, and Garrett 2002, Elbaz et al. 2002).

We applied the Elbaz et al. (2002) prescriptions to estimate the bolometric IR
luminosity of our sources.
At a redshift of $z>0.8$ the LW3 band corresponds to the LW2 rest-frame, 
while for $z<0.8$ the central wavelength of LW3 falls into the IRAS 12 $\mu$m filter. 
We therefore assumed that the luminosities computed from the LW3 fluxes of the sources 
in our sample originate from photons emitted close to the LW2 and IRAS12 wavebands, 
and then used the Elbaz et al. relations:
\begin{equation}
\begin{array}{rcl}
L_{IR} & = &4.78^{+2.37}_{-1.59}\times \left(\nu L_\nu[6.75\ \mu m]\right)^{0.998}\\
&&\textrm{if }\nu L_\nu[6.75\ \mu m]<5\times10^9L_\odot \\
&=&4.37^{+2.35}_{-2.13}\times 10^{-6}\times \left(\nu L_\nu[6.75\ \mu m]\right)^{1.62}\\
&&\textrm{if }\nu L_\nu[6.75\ \mu m]\ge5\times10^9L_\odot , \\
\end{array}
\label{eq3}
\end{equation}
\begin{equation}
\nu L_\nu[15\ \mu m] =0.042\times \left(\nu L_\nu[12\ \mu m]\right)^{1.12},
\end{equation}
\begin{equation}
L_{IR} \sim 11.1 \cdot \nu L_\nu [15\mu m].
\label{eq4}
\end{equation}

From $L_{IR}$ we then computed the SFRs adopting the Kennicutt's (1998) calibration:
\begin{equation}
\frac{SFR}{M_\odot yr^{-1}}=1.72\cdot 10^{-10}\ L_{IR}\ [L_\odot].
\label{eq5}
\end{equation}

The results are reported in the fourth column of Table \ref{tab:sfr_masses_ir}.
Figure \ref{fig:sfr_z} shows a plot of these estimated SFRs as a function of redshift.
The shaded area represents the unobservable region given by our sensitivity limit 
of 95 $\mu$Jy imposed by the sample selection. Error bars have been computed on the
basis of the 1$\sigma$ uncertainties on the LW3 measurments only, while we did not
take into account the scatters in the Elbaz's and Kennicutt's relations.
The plot shows the effect of Malmquist bias induced by the
15 $\mu$m flux limit of the sample, high redshift objects being 
detectable only if their luminosity, and SFR, are large enough.

We compare in Figures A.2 and A.3 the observed spectral energy
distributions of our sample sources with the empirical SED of M82 
scaled to fit the measured LW3 flux (thick continuous line). 
The M82 SED has been taken partly from Silva et al. (1998), partly 
from the observed ISOCAM--CVF low-resolution spectrum between 5 and 
18 $\mu$m by \cite{schreiber2001}, in order to get a proper representation
of the PAH spectrum which is critical for interpreting the LW3 fluxes. 
In the Elbaz's et al. analysis this prototypical starburst falls
close to the barycenter of the L$_{FIR}$ to L$_{15}$ correlation. 
The SFR values appearing in Figs. A.2 and A.3 
refer to these fits, assuming for M82 a $SFR\simeq 6\,M_\odot yr^{-1}$.

We report in Fig. \ref{five} our estimated dependence of the SFR on the 15 $\mu m$ 
flux as measured in the ISOCAM LW3 filter, as a function of redshift, based on the 
M82 SED. This takes into account in detail the effects 
of the filter transmission function and the complex mid-IR spectrum in the 
K-correction factor.
As discussed in Elbaz et al. (2002), values of the SFR based
on the M82 template tend to be lower by $\sim 30-50\%$ than those
based on the more sophisticated analysis based on Eqs. \ref{eq3} to \ref{eq5}.

\subsection{Estimating the baryonic masses}\label{par:bar_mass}

A second fundamental parameter characterizing the sources of the IR background, that
we identified in the faint ISO galaxies, is their mass in stars.
This integral of the past stellar formation activity in the galaxy provides a 
logical complement to the determination of the instantaneous rate of star 
formation.
To measure galaxy masses in the HDF South, we have perfomed spatially-resolved
medium-resolution spectroscopy with ISAAC of few selected targets.  
However, the long time integrations required have limited such dynamical 
estimates to 4 galaxies (see Rigopoulou et al. 2002).

\begin{figure}[!ht]
\centering
\includegraphics[height=0.45\textwidth]{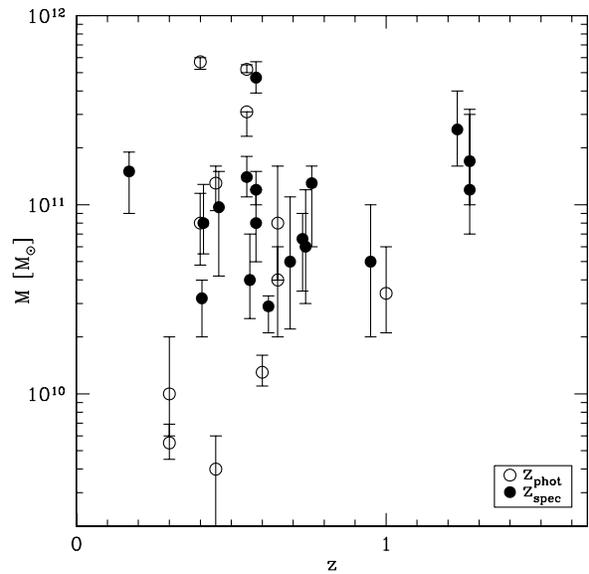}
\caption{Dependence of the masses estimated for the HDF--S 15 $\mu$m sources on
redshift $z$. Solid circles correspond to objects with spectroscopic and open circles
with photometric redshifts.}
\label{fig:mass_z}
\end{figure}

A powerful alternative to the time-expensive spectroscopic investigations makes 
use of observations of the galaxy near-IR SEDs and their moderate dependence 
on the age of the contributing stars (e.g. Franceschini \& Lonsdale, 2002). 
This is due to the fact that in a typical galaxy the stellar mass is dominated by 
low-mass stars, with evolutionary timescales of the order of the Hubble time. 
As discussed by several authors (Lancon et al. 1999, Origlia \& Oliva 2000),
these moderate-mass stars during the cool giant phase largely contribute
to the near-IR SEDs of galaxies.      However, dusty starbursts show occasional 
evidence in the starburst regions for the presence of younger red supergiants as shown
by the pronounced [CO] 2.3 $\mu$m absorption (e.g. Foerster-Schreiber et al. 2001).
Obviously, a contribution by such young massive stars, even heavily extinguished,
would substantially decrease the M/L in the near-IR, hence affecting the stellar
mass estimate.  

We have analysed the UV-optical-NIR data on our sample sources by means of a spectral 
synthesis code that we obtained from a modification of that described in
Poggianti, Bressan \& Franceschini (2001) 
and specifically devised to model dusty starbursting galaxies. 
The integrated model spectrum has been generated as a combination of 10 single stellar 
populations (SSP) of different ages: the youngest generations ($10^6$, $3\ 10^6$, $8\ 
10^6$, $10^7$ yr) and the intermediate ($5\ 10^7$, $10^8$, $3\ 10^8$, $5\ 10^8$,
$10^9$ yr), while the oldest populations of stars have been modelled as a 
constant star formation rate for $2\ 10^9<t<12\ 10^9$ yrs. 

The composite spectrum of each SSP made use of the Padova isochrones with 
the Pickles (1998) spectra, also complemented with spectra from the Kurucz libraries
(Bressan et al. 2001).        Each stellar generation is born with a Salpeter 
IMF between 0.1 and 100 $M_\odot$ and was assumed to be extinguished by dust in a 
uniform screen following the standard Galactic extinction law 
($R_V=A_V/E[B-V]=3.1$). The extinction value E(B-V) was allowed to vary
from one population to another, and the extinguished SSP spectra
were added up to obtain the galaxy synthetic spectrum. Twenty parameters in total --
i.e. the E(B-V) and the stellar mass for each populations -- are needed to define the 
synthetic spectrum.
Note that such large number of independent stellar populations was used to get the 
best possible description of the observed spectra and conservative estimates
of the uncertainties in the model parameters (namely the total stellar mass).

This completely free-form, non-parametric model was devised to account in the most
general way for bursting and discontinuous star formation histories characteristic of 
starburst galaxies, as well as for more normal and quiescent formation patterns.
It also provides easy implementation of different extinction properties for populations 
at different ages.

We have used this spectral synthesis code to perform an exploratory study of how 
degenerate are fits to the observed UV-optical-near-IR SEDs of IR starbursts against 
variations in the age and extinction of contributing stellar populations (Berta et al. 
2002, in preparation). We have explored the model's parameter space
with the {\sl Adaptive Simulated Annealing} method by Ingber (2000), including 
random-number generators, and using $\chi^2$ as a goodness-of-fit test.
Our simulations have shown that the age-extinction degeneracy seriously
hamper the mass estimate, which may be uncertain by factors up to 5 or more
for some dusty objects.
This is partly because the optical-near-IR spectral continuum by itself leaves 
highly undetermined the contribution of supergiant stars to the near-IR flux.

To better constrain the incidence of young extinguished stellar populations and to
evaluate their contribution to the galaxy's M/L ratio, we have included in our 
spectral fitting procedure also the ISOCAM mid-IR LW3 flux, which is a good measure
of the bolometric emission by young stars (see Sect. \ref{par:sfr}). 
To match the observed mid-IR flux, we have computed the far-IR spectrum associated
with a given solution of our spectral synthesis model from the difference between 
the total non-extinguished and the total extinguished UV-optical-NIR flux, and
assuming that this absorbed energy is re-radiated
in the IR with the spectral shape of M82 (see Figs. A.2 and
A.3).
At the end, our observable set included the UV, optical, near-IR
and mid-IR flux data for all sources, as reported in Figs. A.2 
and A.3. Once a best-fitting solution was found, the total stellar mass 
in the galaxy was computed as the sum of the contributions by all stellar populations.

Our estimated values of the baryonic masses M and their uncertainties, as obtained with 
the above procedure, are reported in Table \ref{tab:sfr_masses_ir} (fifth and sixth 
columns) and plotted against redshift in Fig. \ref{fig:mass_z}. 
We see that, although we can exploit a good sampling of the galaxy SED,
the uncertainties in the photometric mass estimates are still fairly large (typically
a factor $\sim$2).
Given the very extensive exploration of the parameter space that we performed, 
we consider these as conservative estimates of the present uncertainties.
Tighter constraints could soon be obtained from IRAC/SIRTF data between 3 and 10
$\mu$m over large sky areas, particularly from the SIRTF Legacy Programs GOODS 
and SWIRE (Dickinson 2002, Franceschini \& Lonsdale, 2002).

The M values do not show significant correlation with $z$, e.g. compared with 
the strong observed correlation of SFR with $z$ (Fig. \ref{fig:sfr_z}). This
reflects our primary selection not being on mass but on the SFR value, 
through the LW3 flux limit.

\begin{figure*}[!ht]
\centering
\subfigure{
\label{fig:sfr_halpha_ir_notcorr}
\includegraphics[height=0.48\textwidth]{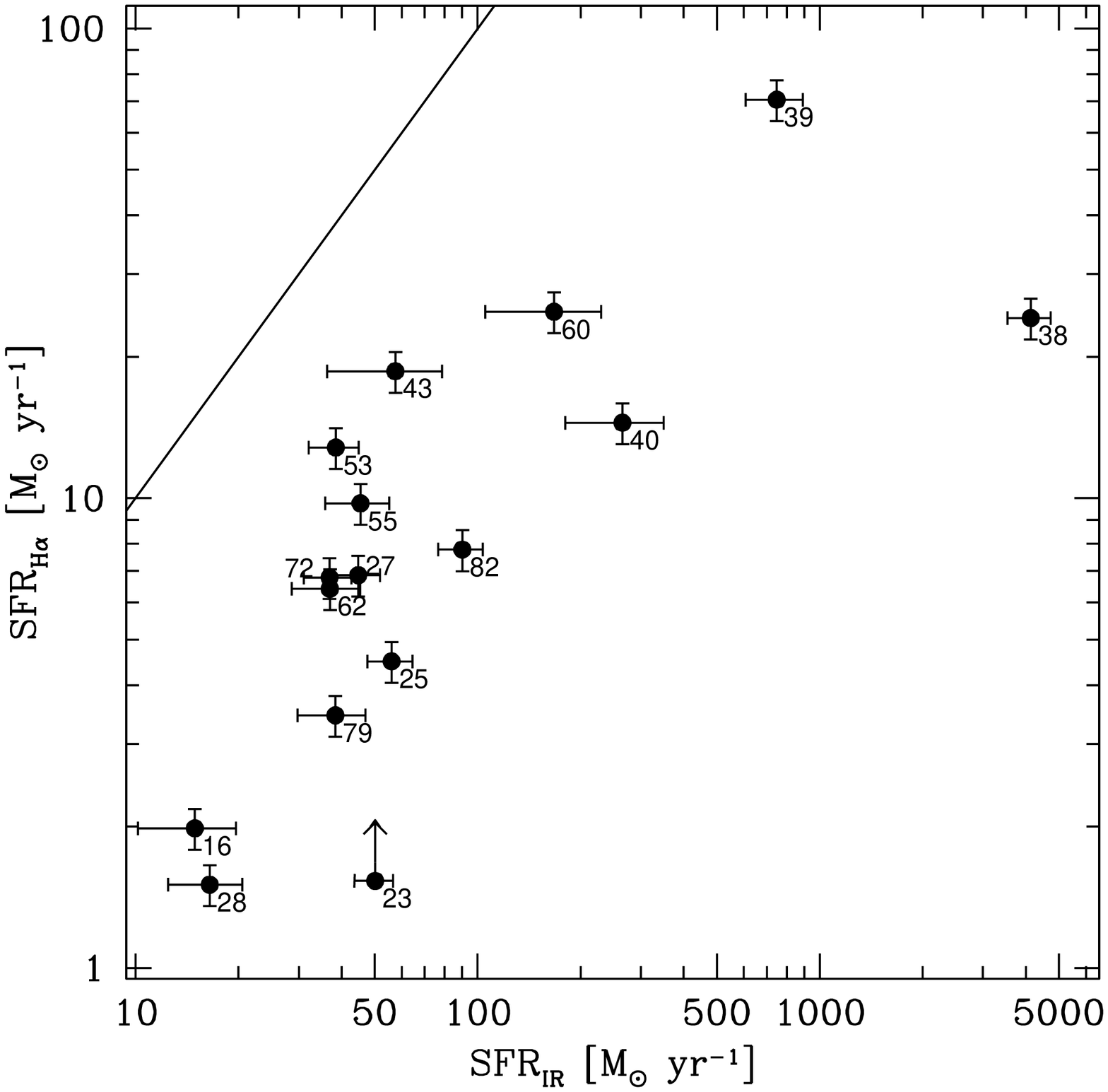}}
\subfigure{
\label{fig:sfr_halpha_ir_corr}
\includegraphics[height=0.48\textwidth]{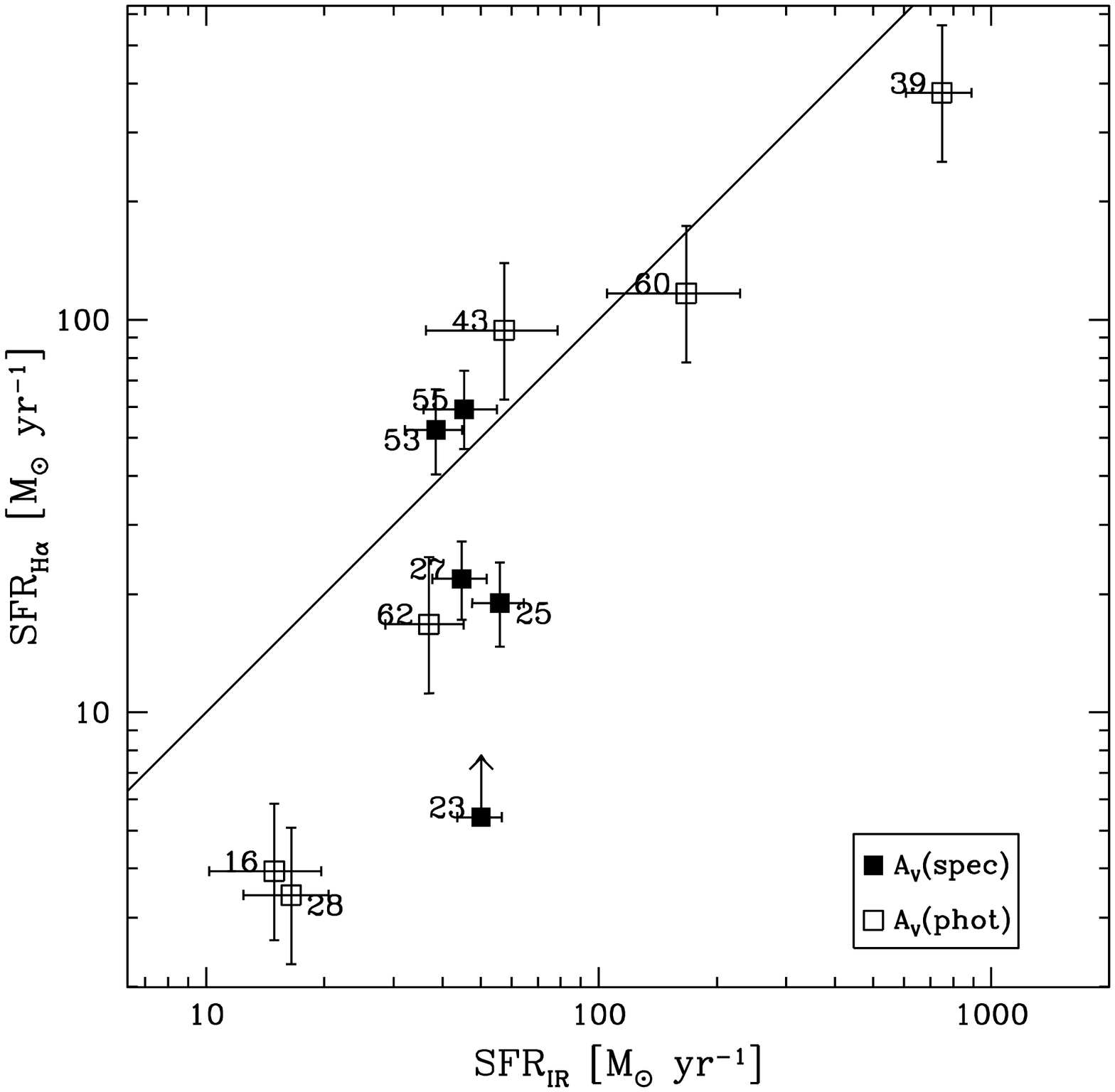}}
\caption{Comparison between estimates of SFR based on the H$_\alpha$
luminosity and those based on SED fitting of the mid-IR flux (as in Table 
\ref{tab:sfr_extinction}).
In panel {\bf (a)} the H$_\alpha$ flux is without extinction correction, 
in panel {\bf (b)} it is corrected for extinction.
Lines indicate the relation $SFR(\textrm{H}\alpha)=SFR(IR)$.
In the right panel open squares refer to $A_V$ estimates based on fits to the
SED, and filled squares to extinction estimated from the Balmer line ratios.
Error bars are based on flux uncertainties (propagating into the extinction
estimate when H$\alpha$ is corrected).}
\label{fig:sfr_halpha_ir}
\end{figure*}

\section{DISCUSSION}

\subsection{Comparison of independent SFR estimators}\label{5.1}

As expected given the nature of these sources selected for their strong mid-IR 
dust emission, the optical estimates of the SFRs based on H$_\alpha$ luminosities 
without extinction corrections provide values very significantly smaller
than those obtained from the mid-IR flux [Fig. \ref{fig:sfr_halpha_ir_notcorr}]. 
Also the H$_\alpha$-based and the IR-based SFR estimates are poorly correlated
with each other.

A large scatter remains even after the H$_\alpha$-based estimates
are corrected for line-of-sight extinction, as illustrated in Figure 
\ref{fig:sfr_halpha_ir_corr}
(open symbols here refer to $A_V$ estimates based on fitting of the optical SEDs, 
filled symbols to estimates based on the Balmer line ratios).


\begin{table}[!ht]
\footnotesize
\centering
\begin{tabular}{ l c c c c r@{.}l }
\hline
\hline
Obj. & $z$ & $\nu$ &  $S_\nu$ & 
$L^1$ & \multicolumn{2}{c}{SFR} \\
\multicolumn{1}{c}{\#} &  & GHz & $\mu$Jy & (1.4 GHz) &\multicolumn{2}{c}{(1.4 GHz)}\\
\hline
n 3 & 1.219 & 8.5  & 56.5 & 154.9 & 1093&00 \\
n 7 & 0.078 & 8.5  & 17.5 & 0.104 & 0&74\\
n 17 & 0.556 & 8.5 & 10.2 & 4.520 & 31&89\\
n 20 & 0.961 & 8.5 & 190.0& 299.4 & 2112&00$^2$\\
n 28 & 0.410 & 8.5 & 26.0 & 5.711 & 40&29 \\
n 32 & 1.275 & 8.5 & 15.1 & 45.84 & 323&40 \\
\hline
s 19 & 1.57 & 4.9 & 163 & 542.0 & 3824&00$^2$\\
s 19 & $''$ & 8.5 & 111 & 547.2 & 3861&00$^2$\\
s 23 & 0.46 & 1.4 & 200 & 14.99 & 105&80\\
s 23 & $''$ & 2.5 & 149 & 14.92 & 105&30\\
s 23 & $''$ & 4.9 & 127 & 17.81 & 125&70\\
s 39 & 1.27 & 1.4 & 109 & 92.70 & 654&00$^2$\\
s 73 & 0.17 & 1.4 & 533 & 4.919 & 34&70\\
s 73 & $''$ & 2.5 & 300 & 4.944 & 34&88\\
\hline
\multicolumn{7}{l}{$^1$ Luminosities at 1.4 GHz in units of $10^{22}$ W
Hz$^{-1}$.}\\
\multicolumn{7}{l}{$^2$ AGN objects: SFR is not a reliable estimate.}
\end{tabular}
\caption{Radio estimate of the SFR of 15 $\mu$m sources in both HDF's and FF's. In
the first column objects named {\em n} are from the HDF--N sample, those with {\em s} in
the HDF--S. The ISOCAM LW3 sources in HDF--N refer to the catalogue by Aussel et al. 
(1999).  Third column specifies the frequency at which the fluxes in the
fourth were measured. Data are from \cite{richards2000} and \cite{mann2002}. }
\label{tab_sfr_radio}
\normalsize
\end{table}

An independent test of the SFR can be inferred from the radio flux: the radio emission 
is also unaffected by dust in the line-of-sight (though it might be sensitive 
in principle to free-free absorption at low radio frequencies for large
column densities of ionized gas, and also sensitive to radio-loud AGN components).
The relationship between the SFR and radio synchrotron emission 
is established by the number of type-II and type-Ib supernova
esplosions per unit time.      \cite{condon1992} finds the relation 
\begin{equation}
\frac{SFR(M>5\,M_\odot)}{M_\odot yr^{_1}} \simeq \frac{L_\nu (1.4\,GHz)}{4\cdot10^{21}
W\,Hz^{-1}}.
\label{radio}
\end{equation}
For our adopted IMF, this becomes
\begin{equation}
\frac{SFR(M>0.1\,M_\odot)}{M_\odot yr^{_1}} \simeq \frac{L_\nu (1.4\,GHz)}{1.2\cdot10^{21}
W\,Hz^{-1}}.
\label{radiotot}
\end{equation}

To improve the statistics, we consider in this and the following sections
source samples selected from deep ISO observations in both the HDF--S and the HDF--N. 
The ISOCAM LW3 datasets in both areas have been reduced in the same way.
As for the HDF--N, the ISOCAM data reduction by
\cite{aussel1999} has detected 77 sources at 15 $\mu$m, 41 of which have 
$S_{15\mu m}>100\,\mu Jy$ and constitute a complete sample over an area
of 25 square arcmin (as for HDF--S). 
HDF--N sources with $S_{15\mu m}<100\,\mu Jy$ do not form a complete sample, 
but they provide an useful extension covering the faint end of the galaxy 
luminosity function. All but 13 of the HDF--N sources have
spectroscopic redshifts, while for the remaining we adopt the usual photometric
estimate.   

\begin{figure}[!ht]
\centering
\includegraphics[height=0.45\textwidth]{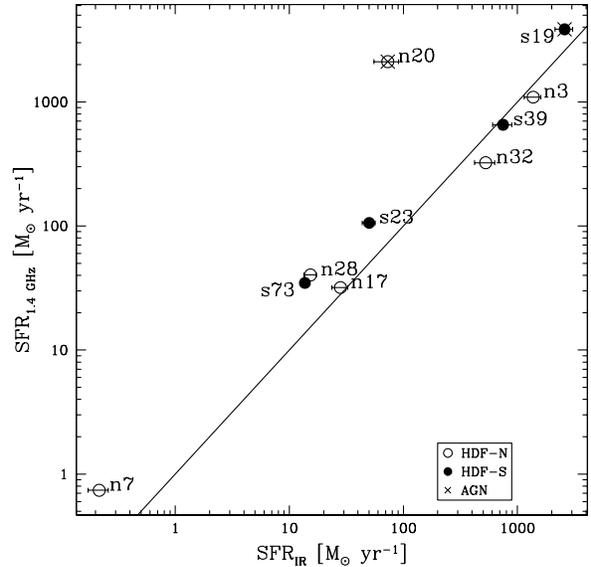}
\caption{Comparison between the estimates of SFR based on the radio flux 
and those inferred from fits to the mid-IR flux. The two methods provide 
consistent values in all cases where the radio is not contaminated by AGN emission
(as is the case for the two crossed sources). Error bars on the IR estimate
are based on the flux uncertainties only.
}
\label{fig:sfr_radio_ir}
\end{figure}

Then using published radio fluxes at 1.4, 2.5, 4.9 and 8.5 GHz for HDF--S galaxies by 
\cite{mann2002} and for HDF--N by \cite{richards1998} and \cite{richards2000}, 
we computed the source luminosities at 1.4 GHz assuming for the radio emission a
power-law with spectral index $\alpha_R$, with $\alpha_R\simeq0.7$ (the mean
value for the HDF--N sample of \cite{richards2000}), except for
sources S23 and S73, for which the spectral index turns out to be 0.5 and 1.0
respectively.

%
%

The SFR values found in this way, and reported in Table \ref{tab_sfr_radio} and 
Fig. \ref{fig:sfr_radio_ir}, are in good agreement with those based on the IR flux.

In both table and figure we report also the data for the sources {\em s19} 
and {\em n20} for which we found evidence for the presence of an AGN ({\em n20}, 
detected by {\sl Chandra} in X--rays, is a recognized type-I AGN, see 
\cite{hornschemeier2001,brandt2001}): a comparison of the 
radio-based SFR value with those reported in Table \ref{tab:sfr_masses_ir}
provides an interesting test of an AGN contribution. 
The good match of SFR values based on the radio and the IR confirms 
the reliability of the latter as a SFR estimator (see also Elbaz et al. 2002,
Garrett 2002).

The rates of SF indicated by our analysis for the faint ISO sources at $z>0.5$
range from few tens to few hundreds solar masses/yr, i.e. a
substantial factor larger than found for faint optically-selected galaxies.

Altogether, our analysis confirms that the mid-IR light is a good tracer of 
the star-formation rate, since it correlates well with the radio and H$_\alpha$ line 
fluxes (Figs. \ref{fig:sfr_halpha_ir}b and \ref{fig:sfr_radio_ir}).
On the other end, even after correcting for dust extinction, SFR estimates
based on the H$_\alpha$ line flux underestimate the intrinsic SFR of luminous
IR galaxies by a factor $\sim 2$, with some large scatter.
We have to consider, however, that large and uncertain correction factors have
been applied due to the poor spatial sampling and dust extinction effects
on the H$_\alpha$ measurement, which might explain it.
Arguments in favour of a large intrinsic scatter bewteen optical line and IR bolometric
fluxes were discussed e.g. in Cram et al. (1998), Poggianti \& Wu (2000),
Rigopoulou et al. (2000), Poggianti et al. (2001), see also Goldader et al. (2002).

This issue will be possibly solved only with the advent of the new-generation 
of optical and near-IR Integral Field Spectrographs on large telescopes 
(e.g. VIMOS and SPIFFI/SINFONI on the ESO VLT, see Thatte et al. 1998),
providing data of high spatial resolution and sensitivity on the optical-UV 
emissions in such morpologically complex systems.

\subsection{Timescales for star formation}\label{par:timescales}

A proper characterization of the evolutionary status of the faint ISO source
population and their relevance for the general process of galaxy formation 
comes from matching the rate of ongoing star-formation SFR
with the mass of already formed stars. Such comparison is essentially independent 
of the stellar IMF (the same scaling factor would apply to both SFR and $M$ 
by changing the IMF).

\begin{figure}[!ht]
\centering
\subfigure[]{
\label{fig:t_z}
\rotatebox{-90}{
\includegraphics[height=0.44\textwidth]{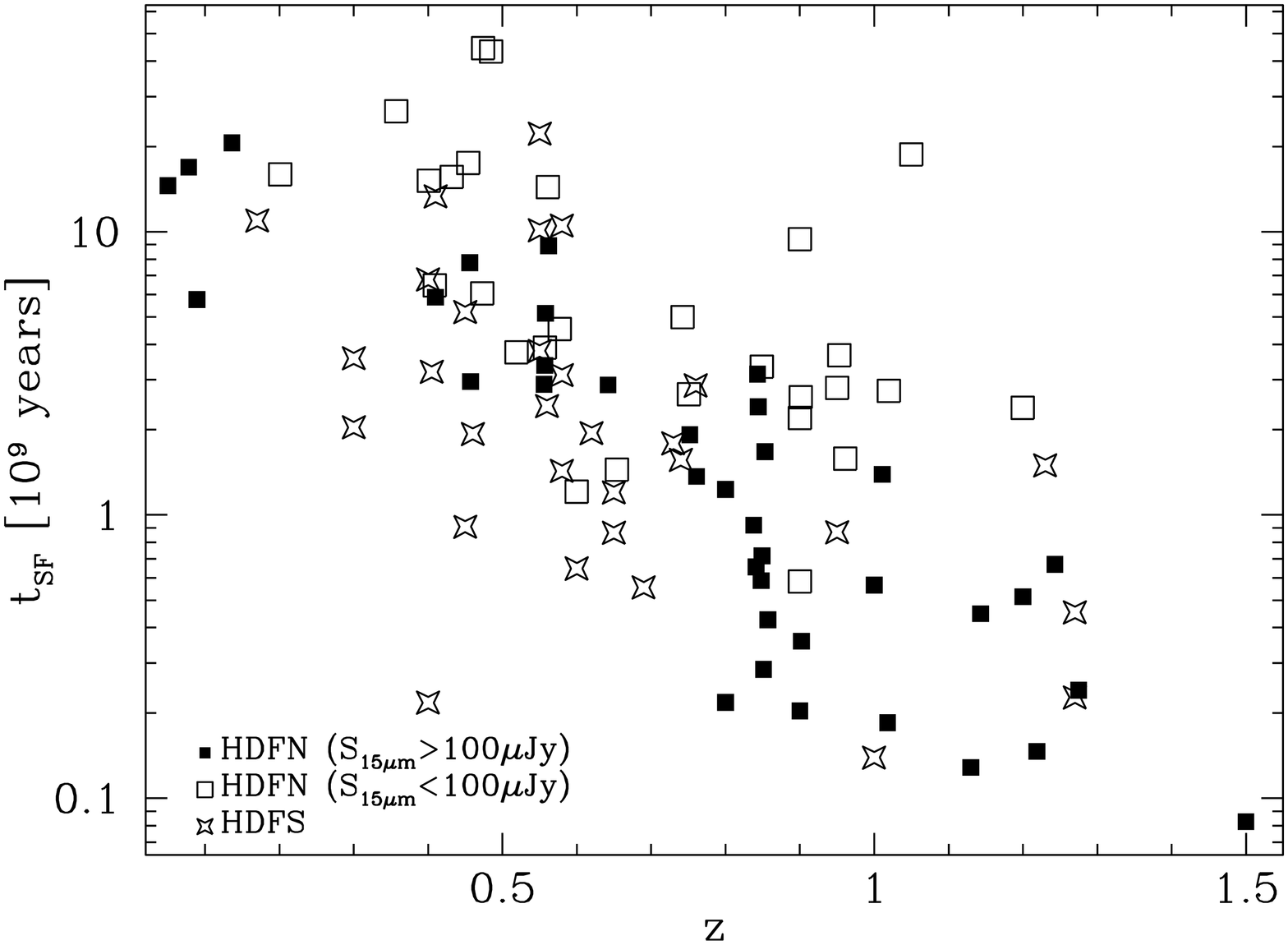}}}
\subfigure[]{
\label{fig:t_mass}
\rotatebox{-90}{
\includegraphics[height=0.44\textwidth]{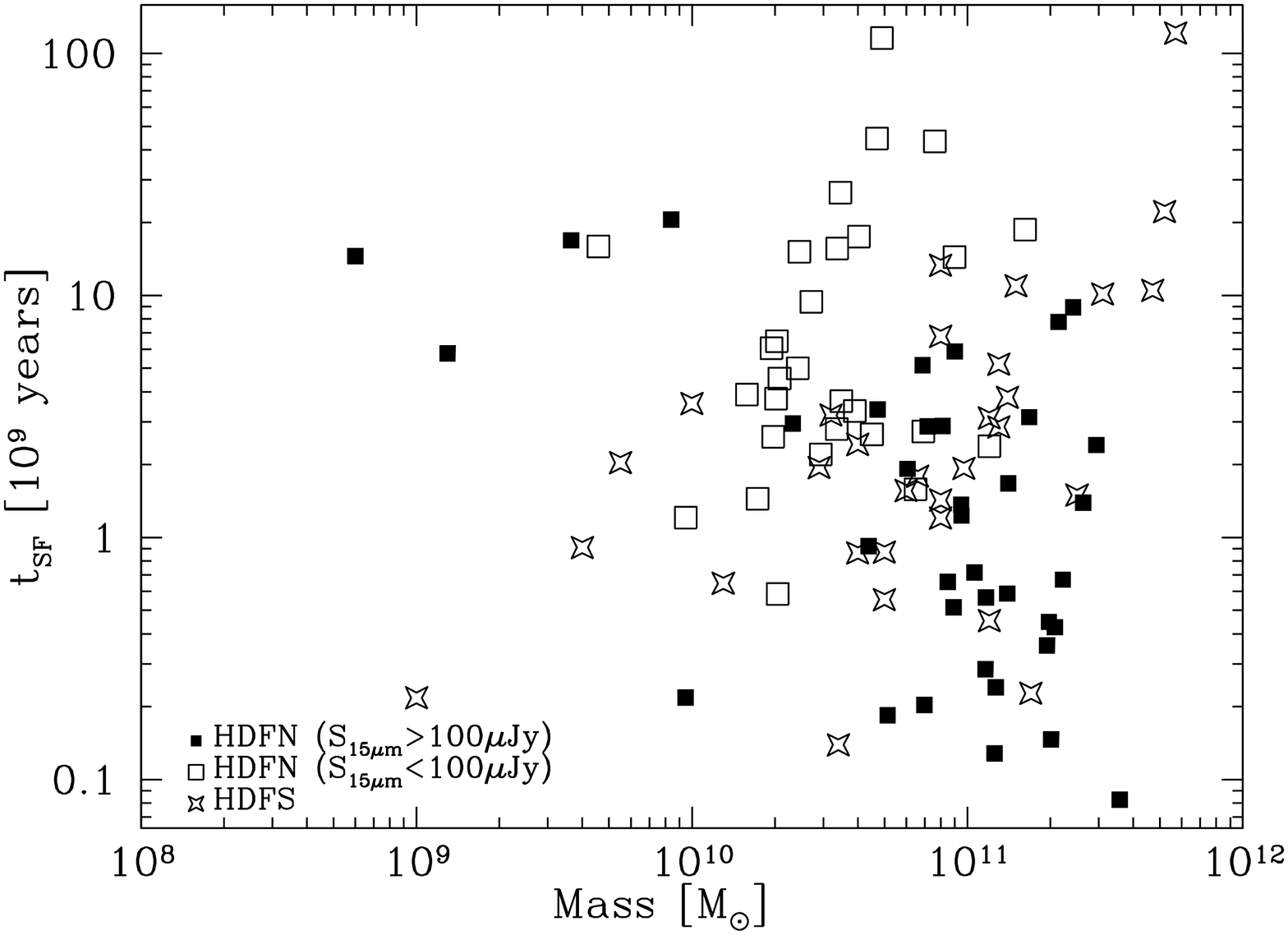}}}
\subfigure[]{
\label{fig:t_sfr}
\rotatebox{-90}{
\includegraphics[height=0.44\textwidth]{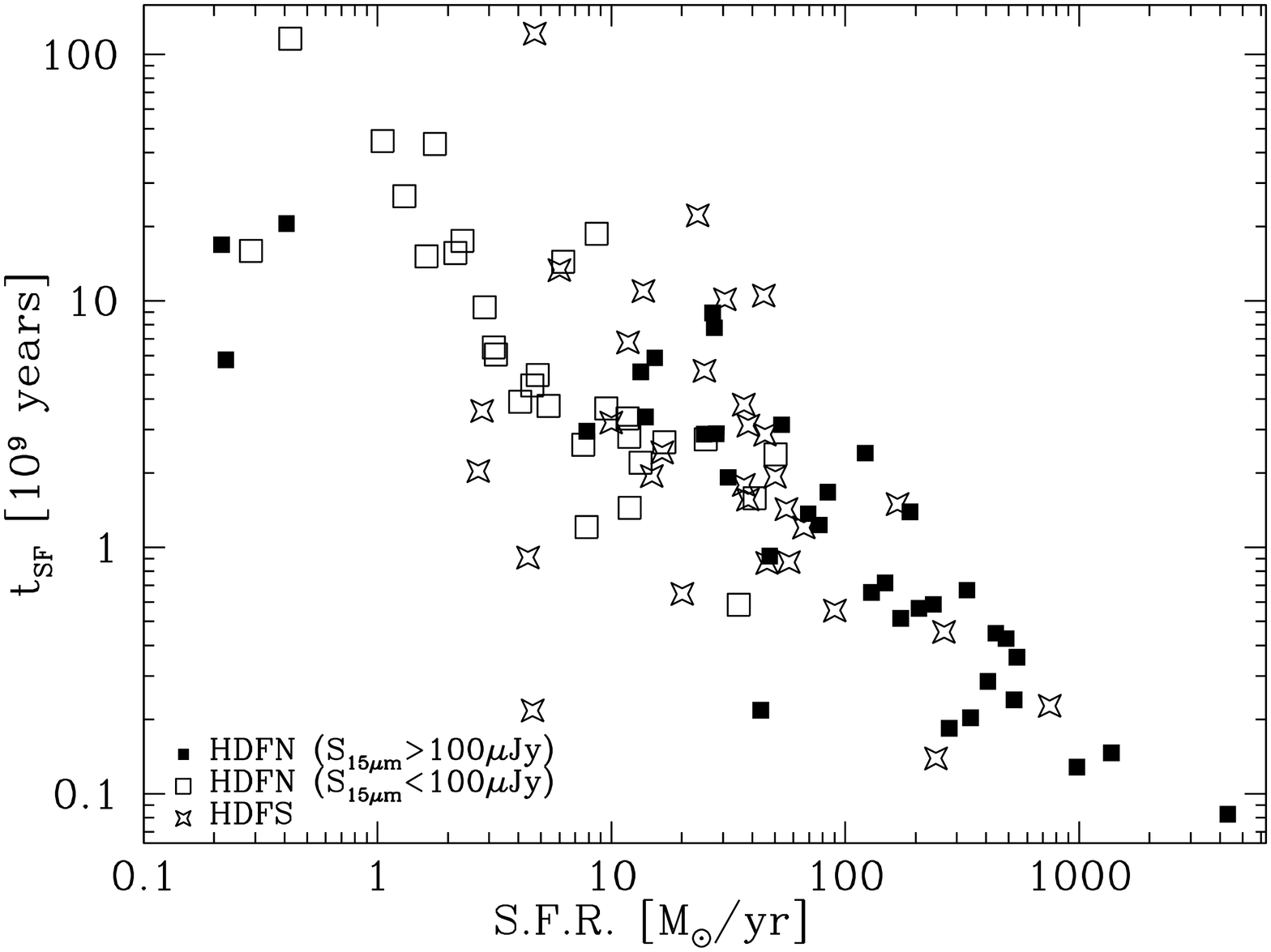}}}
\caption{The timescale of star formation $t_{SF}=M/SFR\ [10^9\ yrs]$ of faint ISO
sources as a function of redshift (panel a), 
mass in stars (panel b) and SFR (panel c).
}
\label{fig:t_zmsfr}
\end{figure}

\begin{figure*}[!ht]
\centering
\subfigure[]{
\label{fig:confronto_ellittiche_bright}
\includegraphics[height=0.48\textwidth]{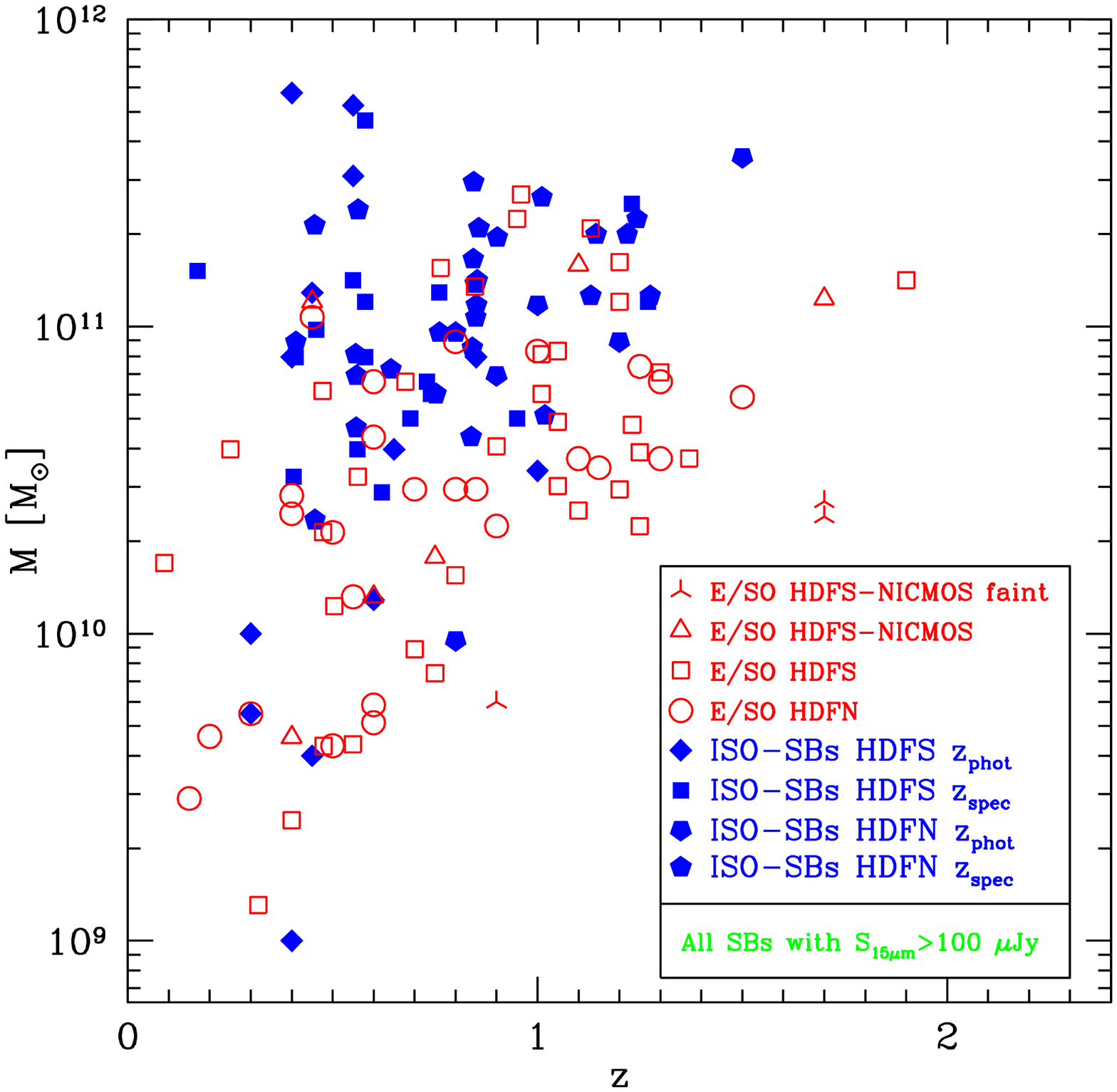}}
\subfigure[]{
\label{fig:fig:confronto_ellittiche_faint}
\includegraphics[height=0.48\textwidth]{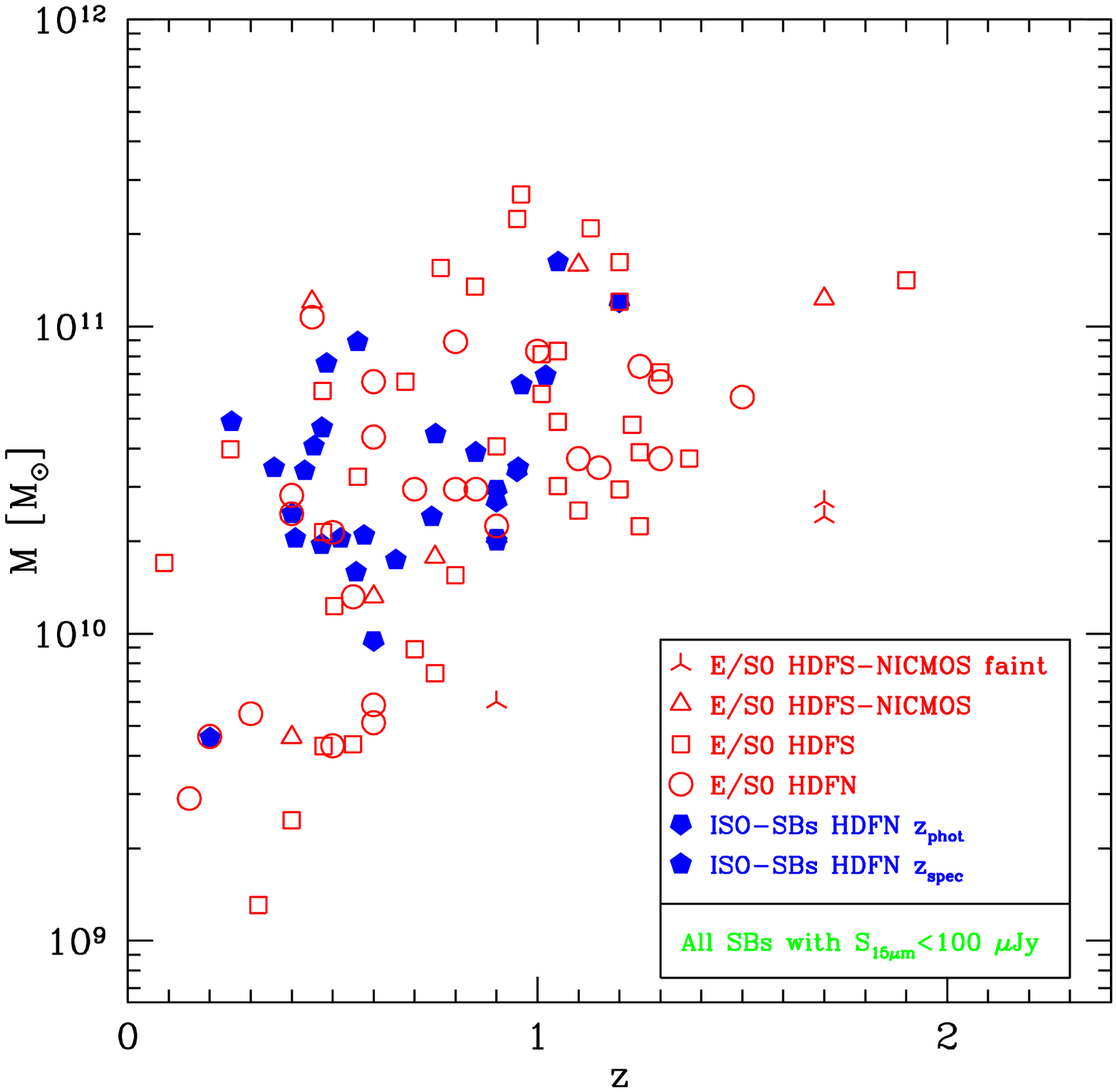}}
\caption{Comparison of the masses determined for the 15 $\mu$m sources in both
HDF, with those of normal elliptical/S0 by \cite{grodig2001}. Panel {\bf (a)}
regards the ISO objects with $S_{15\mu m}>100\mu Jy$, {\bf (b)} those fainter. See text for
comments.}
\label{fig:confronto_ellittiche}
\end{figure*}

We report in Figure \ref{fig:t_z} the ratio of the baryonic mass 
to the SFR against redshift for ISO sources in HDF--S and HDF--N. 
For all these we determined masses and SFRs as discussed in Sect. \ref{par:bar_mass}.
This {\sl activity parameter}, $t_{SF} [yrs]=\frac{M}{SFR}$, is a measure
of the timescale for the formation of stars.
The figure shows that, on top of a very large scatter, there is an 
apparent trend for the SF timescale $t_{SF}$ 
to decrease with $z$ (the Spearman rank coefficient is $\rho=-0.44$ for the
whole sample of 109 sources, with a very significant correlation probability
[$>99\%$]).

On one side, this result indicates that galaxies at $z\simeq 1$ and larger
are more actively forming stars than those in the local universe: a lower
$t_{SF}$ suggests that the ongoing SF is more significantly 
contributing to the observed stellar mass, and a larger fraction of stars
are being formed during the current SF event.
Of course, we expect a selection effect to play, 
if we consider that at the higher redshifts only the IR brightest galaxies can be
detected in the flux-limited samples, while the selection is less directly
influenced by the stellar mass.
We expect that our main selection bias should operate in preventing detection 
of sources in the upper right corner of Fig. \ref{fig:t_z}: less active galaxies, 
those with the highest $t_{SF}$ at the higher redshifts, are not detected 
at 15 $\mu$m due to our limited sensitivity. 

On the contrary, we do not expect that sources are missed in the other corner at 
lower $t_{SF}$ (lower left side in the figure). One effect to consider here is
the low redshift of the sources and the small sampled volume: although 
detecting a luminous massive galaxy becomes less likely here, 
luminosities and masses suffer a similar bias, so the net effect on $t_{SF}$ 
should be negligible.

Figures \ref{fig:t_mass} and \ref{fig:t_sfr} plot the star-formation timescale
against the stellar mass and the SFR. The former shows that there is essentially
no dependence of $t_{SF}$ on M.
Some clear segregation is evident in both Figs. \ref{fig:t_mass} and \ref{fig:t_sfr} 
between sources brighter and fainter than $S_{15}=100\ \mu Jy$:
the latter are systematically shifted towards lower values of the mass and SFR,
and to higher values of $t_{SF}$ (if we exclude a few low-redshift and low-mass 
galaxies). Apparently, the fainter 15 $\mu$m sources correspond to a less "active"
class, closer to the quiescent spiral galaxy population.

Altogether, our analysis indicates a trend for a decreased activity of star-formation
(per unit stellar mass) in galaxies at lower redshifts.

\subsection{Comparison with normal galaxies selected in the K band}\label{par:K}

\begin{figure*}[!ht]
\centering
\subfigure[]{
\label{fig:confronto_spirali_bright}
\includegraphics[height=0.48\textwidth]{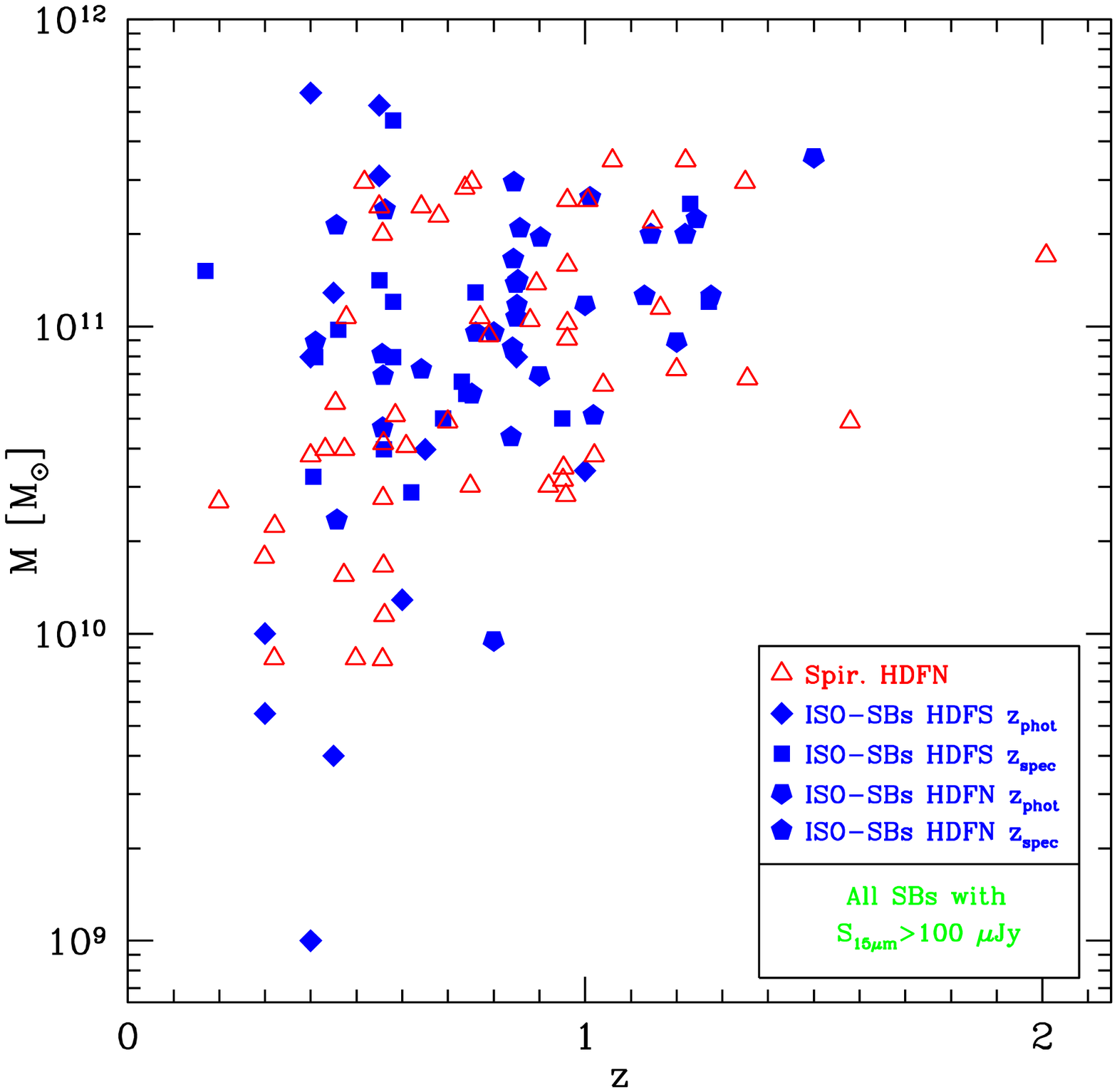}}
\subfigure[]{
\label{fig:fig:confronto_spirali_faint}
\includegraphics[height=0.48\textwidth]{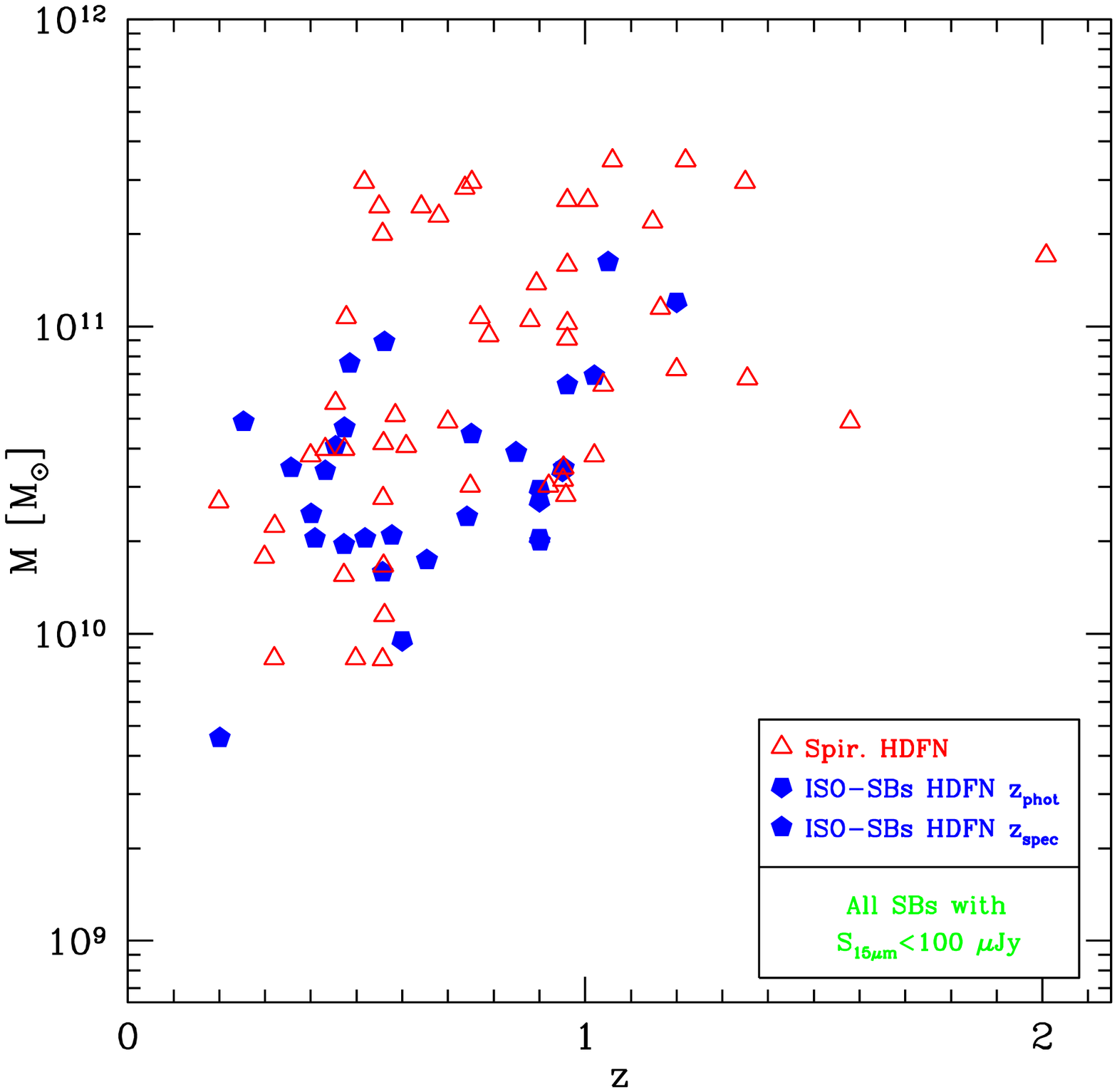}}
\caption{Comparison between the baryonic masses in stars for the 
counterparts to the LW3 sources (filled symbols) and 
those of normal spirals and irregulars in \cite{grodig2000} (open symbols). 
Panel {\bf (a)}: ISO sources with $S_{15\mu m}>100\mu Jy$.
Panel {\bf (b)}: sources with $S_{15\mu m}<100\mu Jy$. 
The meaning of the different symbols are mentioned in the figure insets.
}
\label{fig:confronto_spirali}
\end{figure*}

We gain further insight into the nature of the 15$\mu$m-selected population 
from a match with galaxy samples selected in the optical or near-IR.
We compare here the baryonic masses, while we do not consider 
the SFRs whose estimate based on optical data may be quite uncertain.
Figures \ref{fig:confronto_ellittiche} and \ref{fig:confronto_spirali}
compare the distributions of stellar masses versus redshift for the IR-selected 
sources with those of K-band selected galaxies with morphological classification 
in the HDF--N and HDF--S (filled symbols refer to the ISO galaxies, 
open symbols to the K-selected galaxies).
In Fig. \ref{fig:confronto_ellittiche} the reference sample are 69 
morphologically-classified E/S0 galaxies with
$K<20.15$ in the WFPC-2 HDF--N and HDF--S (plus a few NICMOS--HDFS sources), 
over a total area of 11.7 $arcmin^2$ \citep{grodig2001}. 

The comparison in Fig. \ref{fig:confronto_spirali} is performed 
with a sample of 52 morphologically classified spiral/irregulars with $K<20.47$ 
in the 5.7 $arcmin^2$ WFPC-2 HDF--N field \citep{grodig2000}.
For all datasets, the baryonic masses are consistently estimated from fits to 
the optical/near-IR SEDs, as discussed in
Sect. \ref{par:bar_mass} (previous results by Rodighiero et al.
have been corrected for consistency with our assumed cosmology).
Panels (a) and (b) in both figures concern the bright ($S_{15\mu m}>100\mu Jy$) 
and faint ($S_{15\mu m}<100\mu Jy$) 15 $\mu$m subsamples respectively.
These figures provide evidence that the IR-selected sources include
a population of rather massive galaxies when compared with those selected in K,
(particularly if we consider that, for reasons mentioned in Sect. \ref{par:bar_mass},
the K-flux limit is expected to select the most massive objects). 
This is particularly apparent among the IR brighter sub-sample
(see panels [a]), whose mass distribution shows a tendency 
to occupy a region of high mass values compared with K-band selected E/S0 galaxies.
Fainter 15 $\mu$m sources (panels [b]) appear to be hosted by more moderately massive 
systems.

Note that, in any case, this comparison of IR-selected and K-band selected galaxies
has to be taken with some care due to the different sizes of the fields covered by 
the ISO surveys ($\sim 50\ arcmin^2$) and by the near-IR samples (5 to 10 $arcmin^2$). 
These latter are particularly subject to variance effects due to the small sampled
volumes. The ISO samples themselves show evidence of conspicuous structure: 
two overdensities or clusters are 
evident in Figs. \ref{fig:confronto_ellittiche} and \ref{fig:confronto_spirali}
at $z\simeq 0.58$ for LW3 sources in HDF--S and at $z\simeq 0.84$ in the
HDF--N: associated with these we can notice galactic mass
distributions spread towards high values.
Mann et al. (2002) report a detailed discussion of the cosmic variance and 
sampling effects in the HDF--S.

\subsection{Nature of the IR-selected galaxy population}\label{par:nature}

It is evident from the previous discussion that the ISO surveys above 100 $\mu Jy$ 
select a population of massive galaxies which are associated with active 
sites of star formation.
The WFPC-2 images show clear evidence for interactions and mergers in $\sim$50\%
of these galaxies (Berta, 2000; see also Fig. A.2).
The redshift survey in the HDF--N by Cohen et al. (1999) has also revealed 
that the LW3 sources are almost invariably associated with peaks in the sample 
redshift distribution, which may be interpreted as either "walls" in the galaxy 
large scale structure, or galaxy groups. In such environment galaxy interactions
are maximally efficient.

How much of these large stellar masses are formed during the ongoing
event of star formation identified by the excess mid-IR emission?
Are the currently observed episodes of star formation responsible for the
bulk of the stellar content?
Hints can be inferred from the star-formation timescales reported in Fig. 
\ref{fig:t_zmsfr}. Values of the {\sl activity parameter} $t_{SF}$ here show a 
large scatter and rather uniform distribution between those characteristic of 
inactive galaxies ($t_{SF}\sim 10^{10} yrs$) and those of very active starbursts
with $t_{SF}< 10^{9} yrs$. 
The observed median value t$_{SF}\sim 1$ Gyr, compared with typical starburst 
timescales of $\sim$0.1 Gyr, implies that {\sl a single starburst event may explain
only a fraction of the whole stellar content. A few to several such episodes are 
then required
to build up the observed large galactic masses, as part of a protracted SF history 
made of a sequence of starbursting episodes on top of a lower-level secular SF.
}

Another interesting fact illustrated in Fig. \ref{fig:t_z} 
is that the parameter $t_{SF}$ expressing the "activity stage" of the ISO sources is
rather uniformly distributed between values corresponding to {\sl maximal} activity
($t_{SF}\leq 10^{9} yrs$) and those corresponding to almost inactive systems
($t_{SF}\geq 10^{10} yrs$). This implies that in between the levels of {\sl maximal} 
and {\sl minimal} activity there is a continuity of intermediate stages. 

In conclusion, the mid-IR flux efficiently discriminates distant IR
galaxies as for not only their "activity level" ($t_{SF}^{-1}$), but also
for the absolute values of the stellar mass and SFR: the brighter 
($S_{15}>100\ \mu Jy$) IR-selected sources appear to be more extreme in any sense,
typically more massive, more luminous, and with higher SFR.
This star formation is likely triggered by strong dynamical 
interactions and mergers among the brightest members of galaxy groups.
High SFRs and already large M values imply that these are presumably the formation 
sites of the most massive current-day galaxies.

\section{Conclusions}\label{par:conclusions}

We have investigated the characters of IR emissions in galaxies over a 
fair cosmic volume between $z\sim 0.2$ and $1.5$. In particular, we have
systematically exploited the mid-IR flux as a best and most reliable tracer of 
star-formation. This study makes use of combined observations with the 
{\sl Infrared Space Observatory} and ESO VLT of faint mid-IR sources
selected with the ISOCAM broad-band filter between 12 and 18 $\mu$m down to
$S_{15}\sim 100\ \mu Jy$ and below. 

Previous analyses by Elbaz et al. (2002) and Franceschini et al. (2001) have 
shown that, given their areal densities, redshifts, and far-IR spectra, 
these faint sources are very likely responsible for an important fraction 
(of the order of 50\%) of the energy density contained in the Cosmic 
IR Background between 10 and 1000 $\mu$m.
In consideration of the fact that the CIRB is the major radiant component in 
the universe after the CMB and that it involves more energy than the UV, optical,
and X-ray backgrounds put together, we expect that these objects have played an 
important role in the process of galaxy formation.

We have reported on low-resolution spectroscopic observations with the 
near-infrared ISAAC and optical FORS spectrographs on VLT for a 
representative and unbiased subsample of 21 objects selected in HDF--S. 
In addition to data on line-emissions and the mid-IR fluxes at 15 and 7 $\mu$m, we 
have made use of the rich variety of photometric observations in the UV-optical and 
near-IR to derive further constraints on the physics of the sources.
To improve the statistics, we have also used in our analysis a sample of
mid-IR sources at similar depths in the HDF--N observed by Aussel et al. (1999).

Our main conclusions are hereby summarized.

\begin{itemize}

\item 
Fairly intense H$_\alpha$+[NII] emission is detected in virtually all the observed
sources. The comparison with the H$_\beta$ and [OII], as well as the SED's analysis, 
indicate typically high extinction values $A_V \sim 1.5-2$ to affect the
line ratios, quite larger than found for local normal spirals. 
While obviously coherent with the IR-selection emphasizing excess dust 
emission, this shows that the intrinsic (de-reddened)
H$_\alpha$ flux is strong in these objects.
Enhanced activity is also proven by the fact that the mid-IR flux is typically
larger by factors $>2-3$ than expected for normal galaxies.

\item 
We have investigated evidence for the presence of Active Galactic Nuclei in
the core sample of 21 sources, as possibly responsible for such enhanced activity, 
by combining all available information:
the broadness of the Balmer lines, the morphology of the optical counterparts
from the WFPC-2 images, the shape of the mid-IR spectrum (from the LW3/LW2 flux
ratio), and the ratio of the radio to IR flux.
We have found un-controversial evidence for nuclear activity in only 2 objects 
(S19, a type-I quasar at z=1.57, and S38, a luminous type-II AGN at z=1.39),
while for two other objects (S39, a very luminous ULIRG at z=1.27,
and S82 at z=0.69) we suspect the presence of AGN contributions.
Although the statistics is poor, this result showing a low AGN fraction $\simeq 
10-20\%$ among the faint IR sources is entirely consistent
with that found by Fadda et al. (2002) based on deep hard X-ray data.

\item 
Then assuming that for the bulk of these objects a starburst is the dominant
energy source, we estimated the rates of star-formation SFR from all available
indicators (H$_\alpha$ line flux, mid-IR and radio emissions). 
We confirm that the mid-IR light is a good tracer of 
the star-formation rate, since it correlates well with the radio and H$_\alpha$ line 
fluxes (Figs. \ref{fig:sfr_halpha_ir}b and \ref{fig:sfr_radio_ir}).
We found typically high values of SFR$\sim 10 - 300\ M_\odot/yr$ for our IR-selected 
galaxies.  On the other hand, even after correcting for dust extinction, SFR estimates
based on the H$_\alpha$ line show a large scatter and some systematic 
offset compared with the intrinsic SFR of luminous IR galaxies.
It still remains to be checked with sensitive IR Integral Field spectrographs 
whether this is due to poor spatial 
sampling by the slit spectrograph, or to an intrinsic depletion of the 
optical emission by large dust extinction in the SF regions of the most active
starbursts.

\item 
We have exploited the excellent coverage of the SED in the optical/near-IR to
estimate the second fundamental parameter, the mass in stars M, by making use
of a newly-devised tool to quantify the M/L of stellar populations with different
ages and extinction. We find
that the faint IR sources with fluxes $S_{15}> 100\ \mu Jy$ are hosted by 
massive galaxies ($M\simeq 10^{11}\ M_\odot$), even if compared 
with those selected in the K band at similar redshifts. Spatially-resolved
near-IR spectroscopy of a few of these galaxies by Rigopoulou et al. (2002) 
supports this result.

\item
By matching these large stellar masses with the observed rates of SF,
we have determined timescales for SF of $t_{SF}\sim 0.2$ to 10 Gyrs. When
compared with the $t_{SF}$ values ($\sim 10^8\ yrs$) typically found for starbursts, 
this implies that the ongoing SF can generate only a fraction
of the stellar content in these galaxies, many of such repeated episodes 
during a protracted SF history being required for the whole galactic build-up.
A trend towards a reduced level of star-formation activity in galaxies at 
decreasing redshift is also apparent in our data.

\end{itemize}

In summary, the 15 $\mu$m selection appears to emphasize sites of enhanced
star formation inside massive galaxies, which are typically the brightest 
members of galaxy groups. 
These sources probably trace evolutionary phases, involving strong dynamical 
interactions and mergers, bringing to the formation of massive 
current-day galaxies.

\begin{acknowledgements}
We thank N. Thatte for support during the ISAAC 2001 run.
This research has been supported by the Italian Space Agency (ASI) and the
European Community RTN Network "POE", under contract HPRN-CT-2000-00138.
\end{acknowledgements}


\appendix
\section{Notes on individual sources}\label{par:sources}


Some properties of individual sources are hereby summarized. Imaging data and
SEDs mentioned here refer to those of Figs. A.2. Images are 
from the HDF South WFPC-2 V-606 band observations.

{\bf S14}. Outside the area covered by WFPC-2, 
the {\sc [Oii]}, H$_\beta$ and {\sc [Oiii]} 
lines have been detected with EMMI at z=0.41. The optical spectrum is quite bluer 
than that of M51, and is fit by the SED of a massive, moderately star-forming 
spiral with SFR=$16.5\,M_\odot yr^{-1}$, as also revealed by the modest
excess of the LW3 15 $\mu$m flux above the M51 spectrum.

{\bf S16}. Also outside the HST fields, it shares very similar properties to
S14, with an H$_\alpha$ detection at z=0.62.

{\bf S19} is an WFPC-2 point-like source, detected also at 4.9 and 8.5 GHz
\cite{mann2002}. Its SED (Fig. A.2) is virtually a flat power-law
from 0.3 to 20 $\mu$m. The ISAAC spectrum reveals a strongly
broadened (FWHM $\sim$ 260 \AA) H$_\alpha$ at z=1.57. The corresponding 
velocity field ($\Delta v > 4600$ km/s) indicates this source to be a type-I
quasar. Two broad emission features are also detected in the FORS spectrum
of Fig. \ref{fig:spettro_s19H_fors1}, corresponding to 
Mg{\sc ii}$\lambda 2798$ and {\sc Ciii]}$\lambda 1908$ at redshift $z \simeq 1.56$.
The observed SED is reproduced by a combination of a starburst template 
in the optical and a type I AGN model. The latter is taken from 
Granato, Danese \& Franceschini (1997) and Franceschini et al. (2002), with best-fit
parameters $R=300$ (ratio between inner and outer tori radii)
and equatorial optical depth $\tau_\nu=50$ mag at 0.3 $\mu$m.

{\bf S20}. Very similar to S14 and S16 in all respects, two lines 
detected by EMMI are {\sc [Oii]} and H$_\beta$ at z=0.39. The overall SED
is quite well fit by the M51 template, and the source looks like a massive but 
relatively normal and quiescent ($SFR=11\,M_\odot yr^{-1}$) spiral galaxy.

{\bf S23} has been observed with NTT/EMMI, FORS2 and ISAAC on VLT
(ISAAC observations are from Rigopoulou et al. 2000). It appears as a triple system in the HST image, but
unfortunately only one of the three nuclei (that in the middle) has been included 
in the spectrograph slits because of the limited orientation capabilities. 
Four emission lines are detected at redshift z=0.46. The central source 
in the triplet, which we identify as that mostly responsible for the IR emission,
is extremely red ($B-K=5.2$).
The LW3 flux shows a strong excess and indicates $SFR=67\, M_\odot
yr^{-1}$. The source is also detected in the radio at 1.4, 2.5 and 4.9 GHz
\citep{mann2002}. 

{\bf S25}, outside the WFPC-2 frame, is associated with a likely interacting 
pair of galaxies on the EIS images. We assumed the optical counterpart to
correspond to the brighter galaxy. Both optical counterparts have been observed
in the high-res ISAAC mode by Rigopoulou et al. (2002), and were found to
counter-rotate. H$_\alpha$ is resolved from {\sc [Nii]}. Optical and near-IR
spectra consistently indicate a redshift of 0.58. 
Our photometrically-estimated baryonic mass is $\sim10^{11}M_\odot$, to be 
compared with the total dynamical mass of $4\ 10^{11}M_\odot$.

{\bf S27} was observed by \cite{rigo2000} and here with FORS2. H$_\alpha$ and
H$_\beta$ detections indicate $z=0.58$. Rigopoulou et al. (2002) report results 
of high resolution ISAAC spectroscopy, providing a huge
dynamical mass of $10^{12}M_\odot$. The optical SED is well fit by the M51
template, and yields to a mass of $4.5\cdot 10^{11} M_\odot$. 
The WFPC-2 images show a prominent bulge and well evident regions of 
star-formation along the spiral arms.
The moderately high LW3 flux suggests, however, that although this is one 
of the most massive spirals known with evidence of ongoing star-formation, 
overall it looks as a relatively quiescent galaxy (see Sect. \ref{par:sfr}).

{\bf S28} is a substantial, relatively blue, spiral at $z=0.56$ 
\citep[H$_\alpha$ from][]{rigo2000}, with a clearly defined and asymmetric 
spiral arm, possibly indicative of an interaction.

\begin{figure}
\centering
\rotatebox{-90}{
\includegraphics[height=0.48\textwidth]{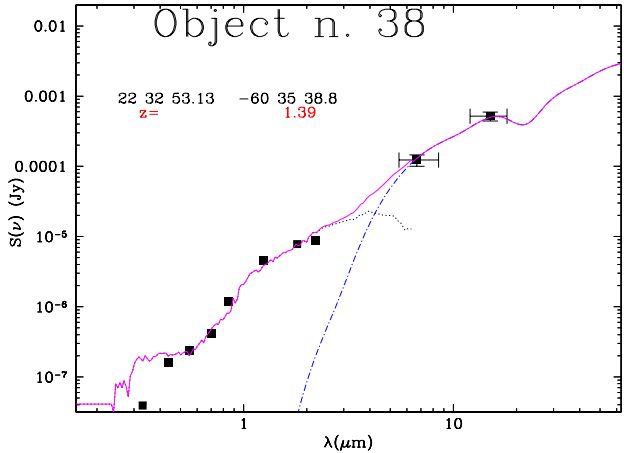}}
\caption{Spectral Energy Distribution of the object {\em s38}.
The optical--NIR datapoints are fitted by normal spiral synthetic spectrum 
obtained with GRASIL (thin solid line, see sect.\ref{par:bar_mass}), while the
ISOCAM mid-IR data are fitted by a type--II AGN spectral template 
(dot--dashed line). The latter assumes obscuration by an edge--on torus whose
ratio between inner and outer radii is 300, and with an optical depth at 0.3
$\mu$m of $\tau_\nu$=30 (see Franceschini et al. 2002).
}
\label{fig:s38_agn}
\end{figure}

{\bf S38} has been observed by \cite{rigo2000}, who detected the resolved
H$_\alpha$ and {\sc [Nii]} lines. The line ratio is inverted
({\sc [Nii]}$\lambda 6583 >$H$_\alpha$), indicating a kind of type-II AGN activity. 
The object is detected at both 15 $\mu$m and in LW2 at 6.7 $\mu$m: the comparatively
large value for the latter also clearly indicates an AGN contribution.
If we assume the IR spectrum of a type-II quasar (Franceschini et al. 2002)
fitting the LW3/LW2 ratio (see Fig. \ref{fig:s38_agn}), the 8 to 1000 $\mu$m
luminosity becomes $L_{IR}=2.72\cdot10^{45}\textrm{erg}\,\textrm{s}^{-1}$.
The very red optical source shows an almost point-like core with a very faint 
diffuse extension (Fig. A.2).   The moderate broadness 
of the H$_\alpha$ and the red optical colors suggest a type-II AGN nature.

{\bf S39} at z=1.27 is the most luminous ULIRG (L$_{IR}=4.4\ 10^{12}L_\odot$) 
among the sources in our sample. It is a radio source detected at 1.4 GHz. 
The WFPC-2 morphology becomes more and more point--like at increasing 
wavelengths, from U to K \citep{mann2002}.
The WFPC-2 image shows a spiral structure with a bright nuclear extended 
emission, a kind of compact bulge.
\cite{rigo2000} detect H$_\alpha$ at $z=1.27$, with marginal evidence for 
broadening (1200 $\pm 450$ Km/s). 
The LW3 flux, assumed it is due to a starburst (but an AGN contribution 
is also possible, see Sect. 4.1),
leads to a very high value of $SFR\simeq 750\,M_\odot yr^{-1}$.

{\bf S40} is a strongly disturbed spiral showing three hot spots, likely
starbursting regions, in the HST image. 
H$_\alpha$ is detected in the ISAAC low-resolution spectrum at $z=1.27$. 
The optical SED is very well fit by the M51 template,
yielding a large baryonic mass of $\simeq 2.2\cdot 10^{11}M_\odot$. The LW3 
flux shows a very large excess, indicative of an actively starforming galaxy with
$SFR\simeq 265\,M_\odot yr^{-1}$.

{\bf S43} is outside the HST field. H$_\alpha$ is detected in the
ISAAC 2000 run at $z=0.95$ (the line was not detected by Rigopoulou et al. [2000]
because of a poor estimate of the photometric redshift). 
The optical spectrum is well fit by M51, with a large mid-IR excess.

{\bf S53} is a beautiful double system with clear signs of interaction. 
Rigopoulou et al. (2000) published the near-IR spectrum of the brighter component. 
In the optical spectra we can deblend two
components of H$_\beta$, at $\Delta \lambda \simeq 30$ \AA\, with EMMI and
$\Delta \lambda \simeq 25$ \AA\, with FORS2. These values yield to a difference
in radial velocity between the two galaxies of about 500 km/s. 
The brighter galaxy shows a well-defined circum-nuclear ring caused by a
previous high-velocity encounter with the smaller galaxy. This latter
shows in turn a faint off-center spot on top of a low-surface brightness
emission.
The redshift measured by \cite{rigo2000} and confirmed here is $z=0.58$.
The optical SED is rather blue, not well matched by the M51 template (the
mass estimate is based on a fit with refined spectrophotometric model).

{\bf S55} is identified with another double interacting optical source, 
including a large spiral and a seemingly spheroid. H$_\alpha$ has been detected by
\cite{rigo2000} in ISAAC low-res and medium-res at $z=0.76$, and H$_\beta$ with
FORS2 observations. The optical-UV SED is resonably well reproduced by M51.
The high resolution spectrum reveals an ``S''--shaped H$_\alpha$ emission, and
allows a dynamical mass estimate consistent with our photometric estimate
(see Table \ref{tab:sfr_masses_ir}).

{\bf S60} is the IR counterpart of a high-redshift spiral galaxy with
a possibly interacting counterpart showing some evidence of interaction (a tail).
It has been observed by \cite{rigo2000} who detected H$_\alpha$ at
$z=1.23$. The overall spectrum is fairly well fit by the M51 template, 
with only a moderate enhancement of the LW3 flux. The galaxy then appears 
to be moderately active in forming stars. However, its photometrically-determined
mass appears to be enormous ($\sim 5\cdot 10^{11} M_\odot$), and similar 
as such to the other system of source S27, apart from the much larger redshift.
Its luminosity and $SFR\simeq 265\,M_\odot yr^{-1}$ are in the ULIRG
regime.

{\bf S62} was observed with ISAAC by Rigopoulou et al. (2000), who detected
H$_\alpha$ at $z=0.73$. The source appears moderately blue in the optical and
quite active in the mid-IR.

{\bf S72} was observed during the ISAAC 2001 run with H$_\alpha$ detected at 
redshift $z=0.55$. The galaxy is an edge-on spiral in an apparently rather 
crowded field, but not much spatial detail is available in the WFPC-2 image. 
The optical SED is quite red, possibly due to inclination. The overall SED is very
well fitted by the M82 template.

{\bf S73} is a very large nearby spiral galaxy with satellites. The H$_\alpha$ line
emission is detected in the FORS2 spectrum at $z=0.17$. The SED has been fitted by us
with the GRASIL library, the mid-IR spectrum showing excess activity
compared with an inactive spiral. The WFPC-2 image shows a spiral with very well 
developed arms and a bar.
Radio emission at 1.4 and 2.5 GHz is reported by \cite{mann2002}.

{\bf S79} H$_\alpha$ line emission at $z=0.74$ was detected in this source during 
the ISAAC 2000 run.  The optical SED matches that of M51, but shows a large excess 
above it in the mid-IR.

{\bf S82} consists of two rather symmetric optical sources, one of which 
almost point-like and the other slightly more extended, but with no much evidence 
of disk-like or otherwise diffuse emission.
Observed with ISAAC in 2000, its spectrum yields H$_\alpha$ at $z=0.69$.
The optical SED is peculiar, with a sharp change in slope in the I band.
The small LW3/LW2 ratio may indicate the presence of an AGN in one or the other
of the optical counterparts.



\end{document}